\documentclass[manuscript,screen]{acmart}
\usepackage{booktabs}
\usepackage{graphicx}
\usepackage{amsmath}

\acmJournal{TOS}
\title{TailGuard-GDS: Tail-Safe Control of GPU Storage Data Movement}

\author{Jiaqi Liu}
\email{liujiaqi242@mails.ucas.ac.cn}
\affiliation{%
  \institution{Institute of Computing Technology, Chinese Academy of Sciences}
  \city{Beijing}
  \country{China}}
\affiliation{%
  \institution{Hangzhou Institute for Advanced Study, University of the Chinese Academy of Sciences}
  \city{Hangzhou}
  \state{Zhejiang}
  \country{China}}

\author{Fan Yang}
\authornote{Corresponding author}
\email{yangfan2020@ict.ac.cn}
\orcid{0009-0004-8087-5903}
\affiliation{%
  \institution{Institute of Computing Technology, Chinese Academy of Sciences}
  \city{Beijing}
  \country{China}}

\author{Xiaohang Shi}
\email{shixiaohang24z@ict.ac.cn}
\affiliation{%
  \institution{Institute of Computing Technology, Chinese Academy of Sciences}
  \city{Beijing}
  \country{China}}

\author{Xianzhe Zhang}
\email{zhangxianzhe25@mails.ucas.ac.cn}
\affiliation{%
  \institution{Institute of Computing Technology, Chinese Academy of Sciences}
  \city{Beijing}
  \country{China}}

\author{Tao Jiang}
\email{jt@ncic.ac.cn}
\affiliation{%
  \institution{Institute of Computing Technology, Chinese Academy of Sciences}
  \city{Beijing}
  \country{China}}

\author{Guojun Yuan}
\email{yuanguojun@ncic.ac.cn}
\affiliation{%
  \institution{Institute of Computing Technology, Chinese Academy of Sciences}
  \city{Beijing}
  \country{China}}

\author{Zhan Wang}
\email{wangzhan@ncic.ac.cn}
\affiliation{%
  \institution{Institute of Computing Technology, Chinese Academy of Sciences}
  \city{Beijing}
  \country{China}}

\author{Guangming Tan}
\email{tgm@ncic.ac.cn}
\affiliation{%
  \institution{Institute of Computing Technology, Chinese Academy of Sciences}
  \city{Beijing}
  \country{China}}

\begin{document}

\begin{abstract}
AI runtimes move state between NVMe storage and GPU memory on critical paths.
GPUDirect Storage (GDS) removes host staging, yet our eight-H100 measurements
show it is not consistently tail-optimal for semantically identical GPU-buffer
operations. The preferred path varies with operation, request size,
concurrency, placement, and resource contention; bandwidth and tail-latency
winners can disagree. Frameworks specify what moves and when, while GPU I/O
stacks expose mechanisms; neither combines phase semantics with measured tail
behavior to choose a delivery plan.

We present \emph{TailGuard-GDS}, a runtime that selects a semantically
equivalent plan per application phase. It profiles plans varying backend,
request shape, and placement, minimizes GPU-visible tail latency under phase
constraints, and revisits decisions as conditions change. Resource reuse keeps
policy lookup and switching at microsecond-scale overhead.

Integrated with PyTorch FSDP and DeepSpeed ZeRO-3, TailGuard reduces FSDP step
p95 by 18.4\% and GPU-idle p95 by 25.3\% versus the strongest fixed path.
Across five eight-GPU ZeRO-3 workloads, step p95 falls 7.9--24.6\% versus fixed
host staging and 28.4--42.6\% versus fixed GDS. Against tuned DeepSpeed AIO,
TailGuard reduces step p95 by 5.6\% and raises estimated throughput by 6.1\%.
\end{abstract}




\maketitle
\section{Introduction}

AI systems increasingly move model and runtime state between local NVMe storage
and GPU memory on execution-critical paths. When model and optimizer state
exceed GPU memory, training systems reload parameter shards from NVMe and
offload inactive state to storage
\cite{zero2020,ren2021zerooffload,rajbhandari2021zeroinfinity,flashneuron2021};
they also periodically write checkpoints
\cite{checkfreq2021,checknrun2022,datastates2024,gemini2023}. Serving systems may
similarly fetch model weights or spilled KV-cache blocks from storage on demand
\cite{flexgen2023}.
When dependent GPU computation cannot proceed until these transfers complete,
GPU-storage latency becomes directly visible to the application. In distributed
execution, an unusually slow transfer on one rank can delay a synchronized
training step, checkpoint phase, or serving iteration.

GPUDirect Storage (GDS) offers a seemingly direct solution. An application
issues cuFile I/O to a GPU-resident buffer, while the GDS data path transfers
data between storage and GPU memory without an application-managed host bounce
buffer \cite{nvidia-gds-overview,nvidia-gds-best-practices}. By eliminating
explicit storage-to-host staging followed by a host-to-GPU copy, GDS can reduce
CPU involvement and data-copy overhead. This design motivates an intuitive
default: whenever a GPU-buffer transfer is eligible for GDS, route it through
GDS. Earlier systems exposed file access to GPU code or integrated storage--GPU
peer-to-peer DMA with the host file system, establishing the broader design
space that GDS now makes deployable through cuFile
\cite{gpufs2013,spin2017}.

We find that this static default is insufficient. On our H100/NVMe platform,
GDS is not consistently tail-optimal for application-visible GPU storage
movement. For semantically equivalent transfers to the same GPU-buffer
endpoint, optimized host staging delivers lower p999 latency across most GPU0
request shapes, and this advantage persists when both paths are constrained to
the same target IOPS. Yet GDS remains preferable for selected write-heavy
workloads and favorable GPU-storage placements. Neither path is uniformly
better: the path with lower tail latency changes with operation type, request
size, concurrency, offered load, and the relative placement of the GPU and
storage devices.

This counterintuitive result arises because eliminating the host copy affects
only one component of end-to-end transfer latency. Application-visible latency
also includes request submission, GPU-buffer registration, file-system and NVMe
service, PCIe data movement, and completion synchronization. GDS and host
staging incur and combine these costs differently. For small or highly
concurrent requests, software and synchronization delays can outweigh the copy
avoided by GDS; for larger transfers, avoiding host staging may become more
valuable. Neither the number of copies nor mean bandwidth alone therefore
predicts the tail latency observed by the application.

To choose the GPU-storage delivery plan that minimizes application-visible
stall for each phase, we present TailGuard-GDS, a runtime controller for GPU
storage movement. TailGuard-GDS operates at the interface between the AI
framework and the storage runtime. The framework reports the current phase,
read or write operation, stream role, and performance requirements.
TailGuard-GDS then chooses a complete delivery plan, including the GDS or
host-staged backend, request size and concurrency, and an available GPU-storage
placement. During an initial calibration step, TailGuard-GDS briefly exercises
candidate plans using transfers representative of the current phase and records
their tail latency and throughput. It selects the lowest-tail plan that
satisfies the phase's throughput or deadline requirements and applies that plan
to recurring transfers. If workload or resource conditions change,
TailGuard-GDS recalibrates the affected phase and switches plans only after
repeated measurements confirm a lower tail latency without violating those
requirements.

TailGuard-GDS exposes GDS and host staging through the same GPU-buffer I/O
interface, allowing the runtime to change delivery plans without modifying
model code. Different phases of the same application may therefore use
different paths: a parameter reload may use host staging, while an offload or
checkpoint phase may use GDS when direct delivery results in a shorter stall.
TailGuard-GDS evaluates each plan by the time until dependent GPU work can
proceed, rather than by storage bandwidth alone. It does not make a single path
choice for the entire job, but chooses between GDS and host staging separately
for each application phase. The key distinction is that TailGuard-GDS controls
how an already requested GPU-buffer transfer is delivered: it jointly selects
the backend, request shape, and placement using phase-specific tail
measurements, while leaving tensor placement and model execution unchanged.

This paper makes three contributions. First, it reveals and characterizes
path-dependent tail latency in semantically equivalent GPU-buffer I/O. Across
controlled variations in operation type, request size, concurrency, offered
load, and GPU-storage placement, neither GDS nor host staging consistently
provides the lower tail. Matched-load comparisons show that these reversals
persist when both paths perform the same amount of work, while topology and
block-layer analyses show that they cannot be explained by storage-device
service time alone. These findings establish GPU-storage delivery as a runtime
decision rather than a one-time system configuration.

Second, it introduces TailGuard-GDS, an application-informed runtime controller
that jointly selects the backend, request size and concurrency, and available
GPU-storage placement for each application phase. TailGuard-GDS uses short
calibration measurements to compare the tail latency and throughput of candidate
plans, selects the lowest-tail plan that satisfies the phase's requirements,
and recalibrates when workload or resource conditions change. All candidate
plans preserve the same GPU-buffer completion interface, allowing TailGuard-GDS
to change data-delivery paths without modifying model logic.

Third, it demonstrates that runtime path selection translates into
application-level improvements. Across multi-rank PyTorch workloads and
end-to-end eight-GPU DeepSpeed ZeRO-3 training, TailGuard-GDS reduces step p95
by 7.9--24.6\% over the strongest fixed delivery path. It also retains an
advantage over a separately tuned DeepSpeed AIO baseline, showing that its
benefits extend beyond isolated I/O benchmarks and fixed-path comparisons.

\section{Background, Motivation, and Requirements}
\label{sec:bg}

\subsection{Storage-to-GPU Data Paths}

\begin{figure}[t]
  \centering
  \includegraphics[width=\columnwidth]{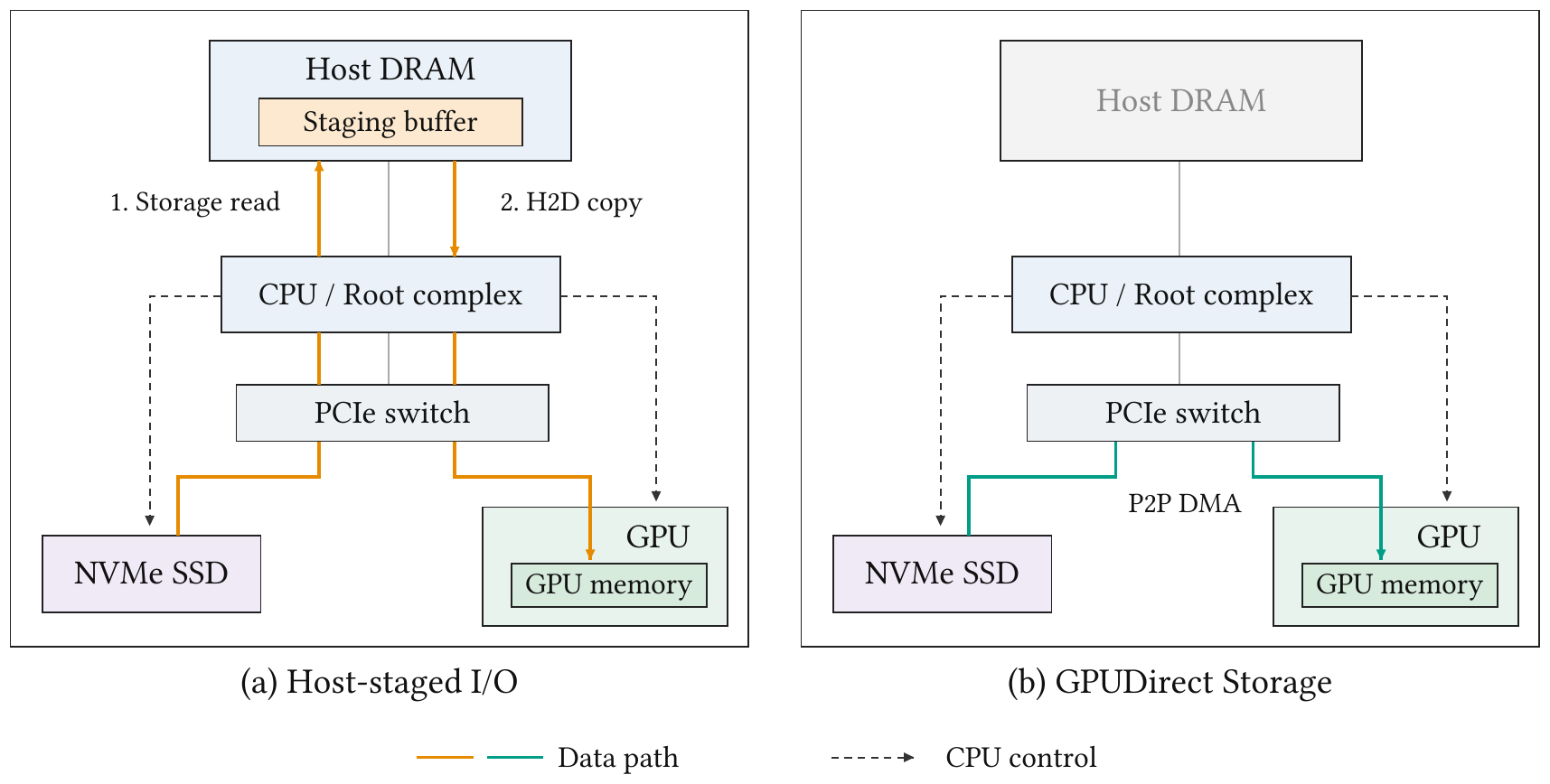}
  \caption{Control and data paths for semantically equivalent GPU-buffer reads.
  (a) Host-staged I/O first transfers the payload from NVMe storage into a
  pinned host-DRAM buffer and then performs a host-to-device copy into GPU
  memory. (b) GDS retains CPU-side I/O submission but transfers the payload
  from NVMe storage to GPU memory through PCIe peer-to-peer DMA, bypassing host
  DRAM. Solid arrows show payload movement, and dashed arrows show CPU control;
  writes reverse the payload direction.}
  \Description{Two side-by-side storage-to-GPU read paths. Host-staged I/O
  moves data from NVMe through pinned host memory and then copies it to GPU
  memory, while GDS moves the payload directly between NVMe and GPU memory over
  PCIe peer-to-peer DMA. Both retain CPU-side control.}
  \label{fig:path-mechanisms}
\end{figure}

GPU applications can move data between files and GPU memory through two common
CPU-issued delivery paths, illustrated in Figure~\ref{fig:path-mechanisms}. In
host-staged I/O (Figure~\ref{fig:path-mechanisms}(a)), the storage read first
DMA-transfers the payload from NVMe through the PCIe and CPU root complexes
into a pinned buffer in host DRAM. A subsequent host-to-device copy sends the
payload back through the PCIe fabric into GPU memory. This path therefore
consumes host-memory bandwidth and GPU copy-engine resources. In GDS
(Figure~\ref{fig:path-mechanisms}(b)), the CPU still submits and controls the I/O
through cuFile and \texttt{nvidia-fs}, but the payload bypasses host DRAM and
moves between NVMe storage and GPU memory through PCIe peer-to-peer DMA
\cite{nvidia-gds-overview,spin2017}. Writes reverse the data arrows. GDS therefore
changes the data path while retaining CPU-side I/O submission.

GDS and host staging implement the same application-visible GPU-buffer
operation. A read completes only after the requested bytes are available in the
target GPU buffer and safe for dependent GPU work. A write completes after the
source bytes have left the GPU buffer and the corresponding I/O operation has
completed under the same write-completion rule for both paths; when durability
is required, the same flush requirement is applied to each path. The two
implementations use the same file, offset, payload, GPU-buffer endpoint, and
completion condition. This common contract enables both a fair performance
comparison and transparent runtime switching between the two paths.

Although GDS and host staging complete the same GPU-buffer operation, they rely
on different system resources. The path with lower tail latency can therefore
change with the operation type, request size, concurrency, GPU-storage
placement, and current resource pressure. The application impact of that
latency also depends on whether the transfer blocks dependent GPU work or
executes as a background stream. These dependencies motivate the tail-oriented
measurements and cost analysis that follow.

\subsection{Why Tail Latency Matters}

Storage tail latency becomes an application stall whenever dependent GPU
computation or rank synchronization must wait for a transfer to complete.
During training, the next compute region may depend on batch staging or the
reload of a parameter shard into GPU memory, while an offload or checkpoint
write can delay a later phase when its completion is required
\cite{rajbhandari2021zeroinfinity,checkfreq2021,checknrun2022,datastates2024}.
In a multi-rank job, the step progresses at
the speed of the slowest rank, so an infrequent slow I/O can become a step-level
straggler \cite{dean2013tail}. Average storage bandwidth may therefore remain
high even while GPUs spend substantial time waiting for storage.

We therefore distinguish path-level evidence from application-level
performance. At the path level, I/O p999 exposes rare delays that average
latency and bandwidth can hide, while achieved bandwidth or IOPS verifies that
the path sustains the required data rate. At the application level, phase stall,
GPU idle time, and synchronized step p95 reveal whether those delays postpone
useful GPU work. TailGuard-GDS selects a lower-tail plan only when it also
satisfies the throughput or deadline requirement of the current phase.

\subsection{Path Choice Is Workload-Dependent}

\begin{figure}[t]
  \centering
  \includegraphics[width=\columnwidth]{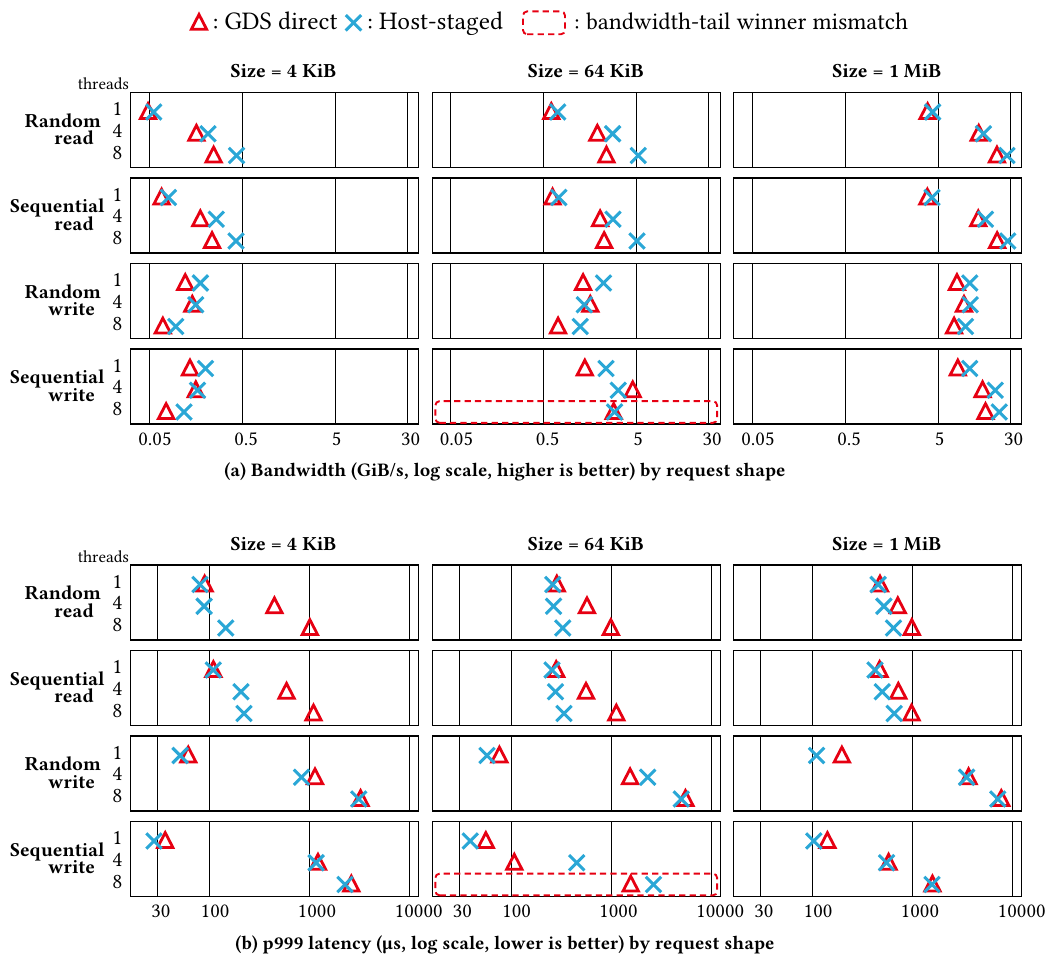}
  \caption{Strict, semantically equivalent comparison across 36 saturated
  workloads. Each paired marker reports GDS direct and host-staged delivery for
  the same operation, request size, and thread count. Panel (a) reports
  bandwidth; panel (b) reports p999 latency. Red dashed boxes identify cases in
  which the bandwidth winner and tail-latency winner differ.}
  \Description{Paired GDS and host-staged results for 36 request shapes. The
  upper panel compares bandwidth and the lower panel compares p999 latency;
  highlighted cases have different bandwidth and tail-latency winners.}
  \label{fig:semantic-equiv}
\end{figure}

Figure~\ref{fig:semantic-equiv} compares semantically equivalent GDS and
host-staged GPU-buffer operations across read and write workloads, access
patterns, request sizes, and concurrency levels on our H100/NVMe platform. Host
staging provides lower p999 latency for most request shapes, but selected write
workloads favor GDS. Moreover, the path with higher bandwidth does not always
have the lower tail latency. Relative to selecting the lower-tail path for each
request shape, always using GDS can increase p999 by as much as 6.92$\times$,
while always using host staging can increase it by as much as 4.20$\times$.
Neither path is therefore tail-optimal across all request shapes.

\begin{figure}[t]
  \centering
  \includegraphics[width=\columnwidth]{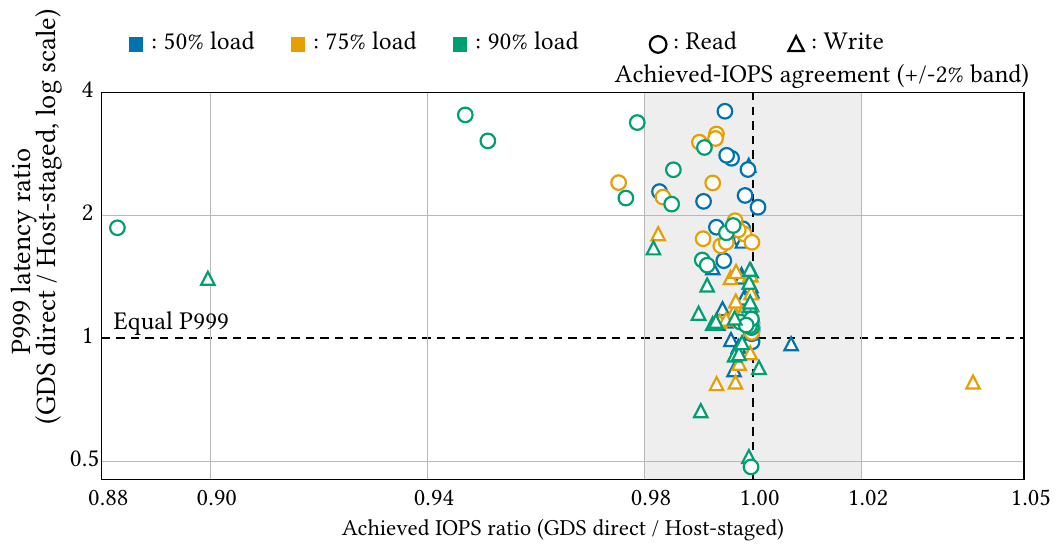}
  \caption{Matched-target comparison at 50\%, 75\%, and 90\% of the lower
  measured capacity. The horizontal coordinate reports the achieved-IOPS
  ratio; the shaded band marks agreement within $\pm2\%$. The vertical
  coordinate reports the GDS/host-staged p999 ratio, and marker shape
  distinguishes reads from writes.}
  \Description{Scatter plot of 108 matched-target comparisons. Most points lie
  near equal achieved IOPS, while the GDS-to-host p999 ratio varies widely and
  frequently exceeds one, indicating a lower host-staged tail.}
  \label{fig:matched-load}
\end{figure}

Saturation measurements can expose the two paths to different request rates,
making it unclear whether their tail-latency difference comes from the delivery
path or simply from unequal load. Figure~\ref{fig:matched-load} therefore gives
GDS and host staging the same target IOPS for each request shape, using rates
derived from the capacity that both paths can sustain. The two paths finish
within 2\% of one another in 100 of 108 cases; eight high-load cases fall
outside this achieved-rate band because one path does not sustain the common
target. Among the 100 in-band cases, host staging has lower p999 in 84 cases,
and the GDS-to-host p999 ratio reaches 3.60$\times$. Thus, the path-dependent
tail difference remains when the paths process comparable achieved loads.

\subsection{Why the Preferred Path Changes}
\label{sec:path-leadership}

Eliminating host staging removes one data-movement stage, but it does not by
itself determine end-to-end completion time. GDS and host staging differ in
request submission, buffer registration, software queueing, PCIe routing, and
completion synchronization before satisfying the same GPU-buffer contract.
Depending on the request shape, placement, and current system state, any of
these components may dominate the observed tail latency. We make these
differences explicit through the following cost decomposition.

\subsubsection{Endpoint-Equivalent Cost Decomposition}

For a request of size $b$, we conceptually decompose the non-overlapped costs
on its completion path as
\begin{align}
L_{\mathrm{gds}}(b) &= A_g + Q_g + T_{s\leftrightarrow g}(b) + S_g, \\
L_{\mathrm{host}}(b) &= A_h + Q_h + T_{s\leftrightarrow h}(b)
                         + C_{h\leftrightarrow g}(b) + S_h .
\end{align}
Here, $A$ includes API submission and resource-management work, including
registration costs that have not been amortized through reuse. $Q$ captures
waiting in file-system, driver, and storage queues; $T$ represents data transfer
between storage and the first destination; $C$ is the explicit CUDA copy
between pinned host memory and GPU memory; and $S$ captures completion and
synchronization visible to the caller. The arrows reverse for writes. These
terms describe the critical path of an individual request; their percentile
values are not additive, and asynchronous execution may overlap work across
different requests.

If we consider only payload movement, GDS appears favorable because the
host-staged path includes both $T_{s\leftrightarrow h}$ and an additional CUDA
copy $C_{h\leftrightarrow g}$. End-to-end tail latency, however, also depends on
submission and resource management $A$, queueing $Q$, and completion
synchronization $S$. A 4~KiB request may spend more time in submission, mapping,
and completion than in transferring payload bytes, whereas a large sequential
request may amortize those fixed costs and increase the value of avoiding host
staging. Reusing a registered GPU buffer reduces $A_g$; changing queue depth
affects $Q_g$ and $Q_h$; and a framework synchronization boundary determines
how much of $S$ becomes visible in step latency. Consequently, the lower-tail
path can change even when the file and hardware remain unchanged.

\subsubsection{Shape, Load, and Placement}

We use \emph{request shape} to describe the operation direction, access pattern,
request size, and concurrency of a transfer. Request size determines how fixed
per-I/O costs are amortized. API submission, address mapping, and completion
synchronization can dominate a small request, whereas a larger request reduces
their relative contribution and places more weight on storage bandwidth, PCIe
transfer, and the CUDA copy in the host-staged path. Access pattern changes the
work performed below the API: sequential offsets permit more contiguous block
requests and more predictable RAID and device service, while random offsets
change request merging, RAID-member access, and service-time variation. Read
and write operations also differ in data direction, file-system processing, and
completion requirements. Finally, concurrency can improve throughput and allow
host I/O to overlap with CUDA copies, but it also increases queue occupancy and
can amplify tail latency. As Figure~\ref{fig:semantic-equiv} shows, these effects
can change the lower-tail path even when the file and GPU remain the same.

Offered load primarily affects the queueing term $Q$. Saturation measurements
reveal the maximum rate that each path can sustain, whereas matched-load
measurements compare their tail latency at the same target IOPS. The persistence
of the latency difference in Figure~\ref{fig:matched-load} shows that it cannot
be attributed solely to unequal request rates.

Placement determines the PCIe and NUMA resources used by each path. With GDS,
data moves between the selected GPU and the devices that hold the requested
file blocks. A GPU and NVMe device under the same PCIe switch can use a shorter
peer path, whereas a transfer across different PCIe branches may traverse
upstream switches or a CPU root complex. Host staging divides the transfer into
storage-to-memory and memory-to-GPU legs; CPU affinity and host-memory placement
can change the locality of these legs, but the path also consumes DRAM bandwidth
and GPU copy-engine resources. Placement is especially important for striped
files because one logical file spans RAID members with different locality
relative to the GPU. In a multi-rail GPUDirect RDMA system, communication can
often be assigned to a NIC close to the GPU, whereas an existing file and its
stripe placement cannot be relocated for each GDS request
\cite{nvidia-gds-best-practices,nvidia-gpudirect-rdma}. TailGuard-GDS therefore
includes available GPU-storage placement in its path-selection context.

\subsubsection{Implication for Runtime Control}

Taken together, these effects show that the lower-tail path depends on the
current application phase, operation, request size and concurrency, GPU-storage
placement, and resource pressure. A runtime controller must combine this
application context with measured path performance, jointly select the backend,
request shape, and available placement, and reconsider the selection when
conditions change. It must also compare candidate plans at the same GPU-buffer
completion boundary: a storage call that returns earlier provides no
application benefit if dependent GPU work can proceed only later. These
observations lead to four requirements for runtime path control.

\subsection{Why Existing Approaches Fall Short}

Existing work addresses neighboring but different decisions in GPU storage
movement. GDS and GPU-native storage systems focus on making a particular
direct path between storage and GPU memory efficient
\cite{gpufs2013,spin2017,flashneuron2021,nvidia-gds-overview,qureshi2023bam,
park2023gids,yuan2025teraio}. They
improve the available data-movement mechanisms, but do not compare semantically
equivalent GDS and host-staged plans according to the stall observed by the
current application phase. AI offload frameworks instead decide which model
state should reside in GPU memory, host memory, or NVMe storage and when that
state should move
\cite{vdnn2016,superneurons2018,capuchin2020,zero2020,ren2021zerooffload,
rajbhandari2021zeroinfinity,flexgen2023}.
Although their I/O engines expose configurable parameters, they do not jointly
select the backend, request shape, and GPU-storage placement for each phase
using application-visible tail latency.

Storage schedulers such as mClock and IOFlow regulate device service rates and
I/O flows to improve isolation or meet storage-level performance objectives
\cite{gulati2010mclock,thereska2013ioflow}. Because they operate below the AI
framework, they do not receive information about whether a transfer blocks
dependent GPU work, belongs to a background stream, or must satisfy a
phase-specific throughput or deadline requirement. Backend-specific tuning can
adjust parameters such as request size, queue depth, and concurrency, but it
optimizes within a selected I/O mechanism rather than deciding among
semantically equivalent GPU-buffer delivery paths. These layers leave a control
gap between application semantics and GPU-storage path behavior.
\emph{TailGuard-GDS fills this gap by combining phase context with measured tail
latency to jointly select the backend, request shape, and available placement
under a common GPU-buffer completion contract.}

\subsection{Requirements for Runtime Path Control}

\noindent\textbf{R1: Preserve GPU-buffer semantics.}
Every candidate plan must preserve the requested file, offset, payload,
GPU-buffer endpoint, and completion condition. A read may complete only after
the requested data is safe for dependent GPU work, while a write must satisfy
the same write-completion and durability requirement under every backend. Path
selection must therefore remain transparent to both model code and application
behavior.

\noindent\textbf{R2: Minimize application-visible stalls.}
Candidate plans must be evaluated by the delay they impose on dependent GPU
work or rank synchronization, rather than by isolated I/O latency or bandwidth
alone. I/O p999 and throughput provide useful calibration signals, but the
selected plan should minimize phase stall or synchronized step tail while
satisfying the phase's data-rate or deadline requirement.

\noindent\textbf{R3: Consider request context and resource placement.}
The lower-tail path can change with the application phase, read or write
operation, request size, concurrency, foreground or background role,
GPU-storage placement, and current resource pressure. The controller must use
this context when comparing complete delivery plans. Selecting only the backend
is insufficient if the resulting plan retains an unsuitable request shape or
placement.

\noindent\textbf{R4: Adapt conservatively to changing conditions.}
The controller must respond to changes in request mix, placement, or resource
contention without switching on noisy measurements. It should remeasure
uncertain candidates and change the active plan only when repeated evidence
shows a tail-latency benefit large enough to cover switching cost while
preserving the phase's requirements. If registration, I/O, or validation fails,
the controller must retain or return to a previously verified path.

\section{TailGuard-GDS Design}
\label{sec:design}

\subsection{Design Overview}

TailGuard-GDS is an in-process control layer invoked at framework storage
boundaries. For each GPU-storage phase, it receives the operation and its
application context, evaluates semantically equivalent delivery plans using
measured tail behavior and phase constraints, and activates the plan expected
to minimize the stall experienced by dependent GPU work. Candidate plans may
differ in backend, request shape, and available GPU-storage placement, but they
expose the same GPU-buffer completion interface to the framework.

These design choices directly address R1--R4. For R1, TailGuard-GDS operates at
framework storage boundaries, where it can observe GPU-buffer completion
semantics, and considers only plans that complete the same logical operation.
For R2, it evaluates each plan by the measured delay imposed on dependent GPU
work while enforcing the phase's throughput or deadline requirement. For R3,
it maintains separate measurements for different phases, operations, request
shapes, GPU-storage placements, stream roles, and backends, preventing
observations from dissimilar contexts from being conflated. For R4, it
remeasures uncertain or stale choices at runtime and switches plans only when
repeated measurements show a benefit sufficient to cover the estimated
switching cost; otherwise, it retains the last verified plan.

\begin{figure}[t]
  \centering
  \includegraphics[width=\columnwidth]{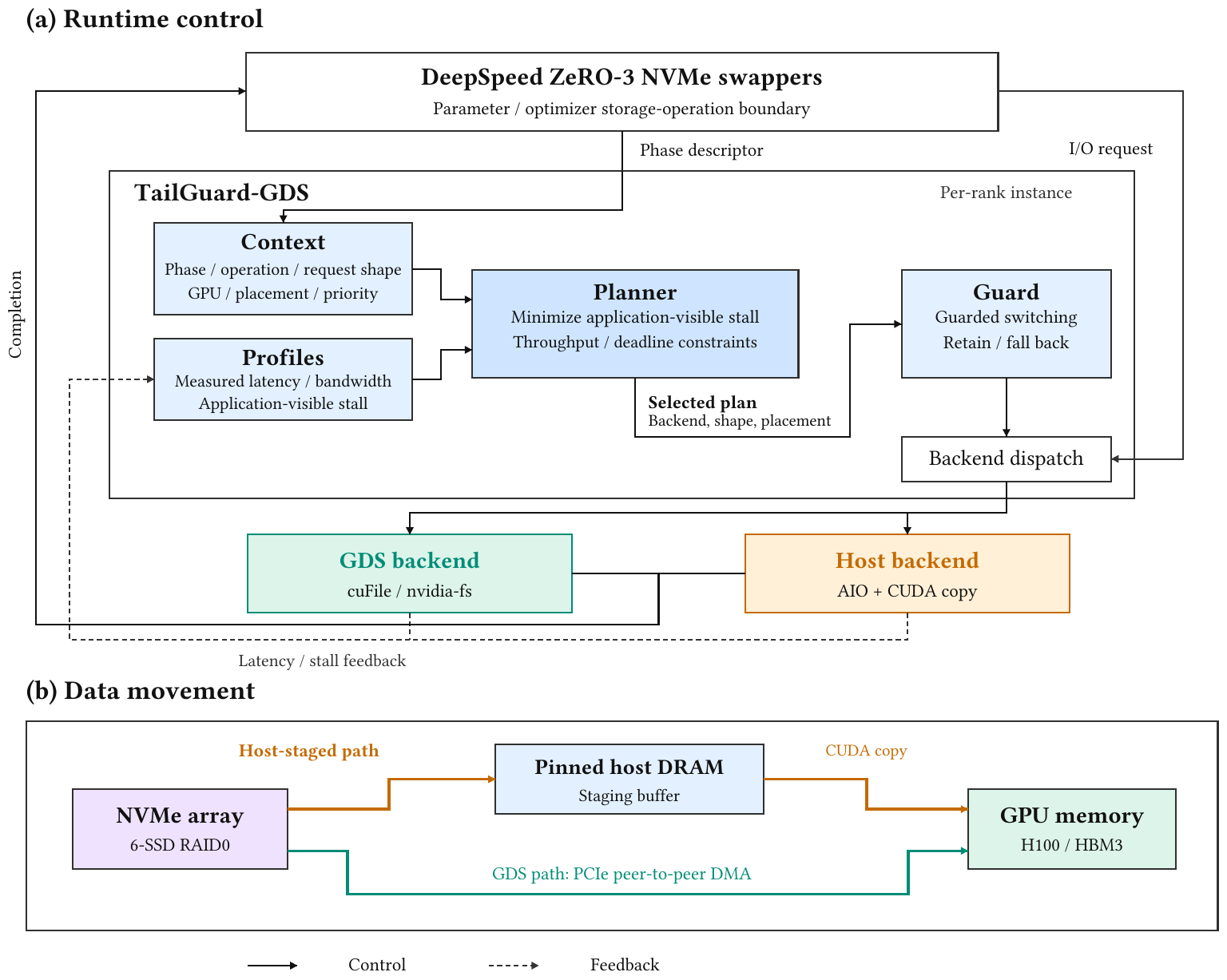}
  \caption{TailGuard-GDS runtime control and data movement. (a) Each rank
  invokes TailGuard at the native DeepSpeed ZeRO-3 parameter and optimizer
  NVMe swapper boundary. Phase and request context, together with measured
  profiles, feed a planner that selects the backend, request shape, and
  placement. A guard validates the change before backend dispatch, while
  completion and latency/stall feedback update the runtime. (b) Both backends
  complete the same GPU-memory operation: GDS uses PCIe peer-to-peer DMA,
  whereas host staging transfers through pinned host DRAM and a CUDA copy.}
  \Description{TailGuard-GDS architecture. Framework phase and request context
  enter a per-rank planner that selects backend, request shape, and placement,
  validates the change, dispatches either GDS or host staging, and updates its
  profiles from completion feedback.}
  \label{fig:runtime-architecture}
\end{figure}

Figure~\ref{fig:runtime-architecture} places the control loop at a concrete
framework boundary. Each rank owns a TailGuard-GDS instance and supplies the
phase descriptor for its storage operation. TailGuard combines that context
with measured profiles, selects a plan, and dispatches the original I/O request
through the admitted backend. Completion returns to the ZeRO-3 swapper at its
existing synchronization point, and the observed latency and application stall
update the profile used by later decisions. The framework-facing completion
contract therefore remains unchanged while TailGuard varies backend, request
shape, and eligible placement.

The design proceeds from representation to runtime control.
Section~\ref{sec:runtime-interface} defines the framework interface and the
representation of candidate delivery plans. Section~\ref{sec:profile-objective}
explains how TailGuard-GDS measures application-visible stall and compares
candidate plans. Section~\ref{sec:calibration-safety} describes calibration,
runtime remeasurement, guarded switching, and fallback.
Section~\ref{sec:mixed-placement} then extends the controller to concurrent
streams and GPU-storage placement choices.

\subsection{Runtime Interface and Decision Tuple}
\label{sec:runtime-interface}

TailGuard-GDS integrates with AI runtimes at their existing storage-operation
boundaries. Before a transfer is issued, a framework hook provides a phase
descriptor containing the operation type, target GPU buffer, stream role and
priority, request shape, and eligible storage placements. TailGuard-GDS uses
this context to select a delivery plan, while the framework continues to
observe completion at its original synchronization point. Parameter reloads,
tensor offloads, batch staging, and checkpoint writes provide suitable
integration boundaries because the framework knows both when each transfer is
initiated and which subsequent GPU computation depends on its completion.

TailGuard-GDS represents each phase invocation by a context tuple
\(w=(p,o,s,g,M,r)\), where \(p\) is the application phase, \(o\) is the
operation, \(s\) is the incoming request shape, \(g\) is the target GPU or GPU
group, \(M\) is the set of eligible storage placements or locality classes, and
\(r\) captures the stream role and priority. For a given context \(w\),
TailGuard-GDS selects a delivery plan \(d=(b,s',\ell)\), where \(b\) is the
backend, \(s'\) is the request shape issued by that backend, and \(\ell\in M\)
is the selected placement. Distinguishing context from plan allows
TailGuard-GDS to compare more than a binary GDS-versus-host choice. For example,
a foreground 4~KiB parameter reload and a background 256~MiB checkpoint write
remain distinct contexts even when they target the same GPU, and each backend
can use a request shape suited to that context.

Regardless of the selected plan, the framework observes the same GPU buffer,
logical file range, and completion point. A GDS plan transfers data directly
between storage and GPU memory, whereas a host-staged plan uses host memory and
an additional CUDA copy. TailGuard-GDS reports read completion only after the
requested data is available to dependent GPU work, and reports write completion
only after the plan satisfies the phase's write-completion and durability
requirement. It may split, merge, or reorder independent requests internally,
provided that the resulting execution preserves the original data and ordering
semantics. TailGuard-GDS can therefore change the delivery plan without
modifying model code or relocating the synchronization point observed by the
framework.

\subsection{Profile and Objective}
\label{sec:profile-objective}

A single performance profile for each backend cannot capture path behavior
because the same backend can exhibit different tails across application
contexts. For example, a large sequential checkpoint write may benefit from GDS
avoiding the host copy, whereas a small fragmented reload may be dominated by
per-request software and registration costs and therefore favor host staging.
TailGuard-GDS instead maintains measurements for each observed context--plan
pair \((w,d)\). In this representation, the request shape in \(w\) summarizes
logical request size, access pattern, batching, queue depth, and concurrency,
while \(d\) identifies the selected backend, issued request shape, and
placement. Each profile entry records the observed latency distribution,
achieved throughput, sample count, and measurement uncertainty. The resulting
profile captures context-specific path behavior rather than imposing a global
ranking between GDS and host staging.

TailGuard-GDS optimizes application-visible stall: the time by which a storage
transfer delays dependent GPU computation or synchronization after any overlap
with computation has been accounted for. This objective differs from raw I/O
tail latency. A background checkpoint write may exhibit a high I/O p999 yet
remain hidden behind computation, whereas a parameter reload on the critical
path can postpone the next compute region and hold other ranks at
synchronization. When the framework exposes the relevant phase boundaries,
TailGuard-GDS measures this stall directly from transfer-dependent wait time,
GPU idle time, or step completion. When direct attribution is unavailable, it
estimates the exposed stall from the request-level tail distribution, the
number of requests on the critical path, their overlap with computation,
whether the slowest request gates a synchronization point, and the interference
observed during execution.

For a context \(w\), TailGuard-GDS selects the delivery plan
\[
  d^*(w)=
  \arg\min_{d\in D(w)}
  \left[
    \widehat{A}(w,d)+\lambda U(w,d)+C(d_0,d)
  \right]
  \quad
  \mathrm{s.t.}\quad
  B(w,d)\ge B_{\min}(w),
\]
where \(D(w)\) contains the semantically valid plans available to context \(w\),
\(\widehat{A}(w,d)\) is the measured or estimated application-visible stall,
\(U(w,d)\) quantifies uncertainty in that estimate, and \(\lambda\) controls how
conservatively uncertain measurements are treated. \(C(d_0,d)\) is the
switching cost amortized over the expected lifetime of the new plan, and
\(B_{\min}(w)\) is the data rate required by the phase. The throughput
requirement may be specified directly by the framework or derived as a fraction
of the highest measured candidate throughput. For a deadline-constrained phase,
\(D(w)\) additionally excludes plans that cannot meet its completion deadline.
These constraints prevent a low-tail but low-throughput plan from extending the
phase, while the uncertainty and switching terms prevent small or unstable
differences from causing unnecessary path changes.

\subsection{Calibration and Safety}
\label{sec:calibration-safety}

Profile entries summarize previous observations rather than guarantee future
performance. Path behavior can change with CPU and host-memory pressure, the
composition of the NVMe request queue, file-system state, and interference from
concurrent workloads. A plan that performed best during initial profiling may
therefore cease to be preferable under different runtime conditions.
TailGuard-GDS uses the profile as a prior and treats an entry as uncertain when
(1) it contains too few samples, (2) its measurements have become stale, or
(3) the current resource conditions differ substantially from those under
which it was collected. TailGuard-GDS verifies such entries before allowing
them to replace the active plan, while reusing well-supported entries observed
under similar conditions directly.

When an uncertain entry could replace the active plan for a recurring or
long-lived context, TailGuard-GDS validates it with bounded calibration probes.
The probes operate on a reserved scratch file located on the same file system
and storage placement as the application data, without reading or modifying
application contents. They reproduce the candidate's operation, backend,
issued request shape, and concurrency, while a fixed request-count or time
budget limits their disruption to the application. TailGuard-GDS uses the
resulting tail-latency and throughput measurements to update the profile and
reduce its uncertainty. If the observations differ substantially from the
existing entry, that entry is marked stale and the current verified plan
remains active until sufficient new evidence is collected.

TailGuard-GDS applies a plan change only at a declared phase boundary and after
outstanding requests issued by the current plan have completed. Three
conditions must then hold: (1) the confidence guard requires the estimated
stall reduction to remain positive after accounting for measurement
uncertainty; (2) the throughput guard requires the candidate to satisfy
\(B_{\min}(w)\); and (3) the switching-cost guard requires the cumulative stall
reduction expected over the lifetime of the new plan to exceed the costs of
draining the current backend, registering buffers, reshaping requests, and
synchronizing the transition. After activation, TailGuard-GDS monitors
application-visible stall, achieved throughput, and backend failures. If the
new plan violates its tail or throughput requirement, or encounters a
registration or I/O failure, TailGuard-GDS returns to the last verified plan,
marks the candidate as uncertain, and imposes a cooldown period before
recalibrating it.

\subsection{Mixed Streams and Placement}
\label{sec:mixed-placement}

An application phase may contain multiple logical storage streams, where each
stream groups requests that share a role, priority, and dependency on GPU
computation. For example, a training step may reload foreground parameter
shards while a previous checkpoint drains in the background; an inference
service may fetch latency-critical state while concurrently evicting or
persisting cache data. Selecting the best plan for each stream in isolation may
not minimize application-visible stall because the resulting plans interact
through shared NVMe queues, PCIe links, host memory bandwidth, CPU resources,
and GPU copy engines. TailGuard-GDS therefore evaluates delivery plans as a
joint assignment across concurrent streams, prioritizing the stream that gates
application progress while enforcing the throughput or deadline requirements
of background streams.

For a mixed phase containing a set of streams \(\mathcal{S}\), each stream \(i\)
provides a context \(w_i\) together with its latency, throughput, deadline, and
placement constraints. A candidate assignment
\(a=(d_i)_{i\in\mathcal{S}}\) selects one admissible delivery plan
\(d_i\in D(w_i)\) for every stream. TailGuard-GDS evaluates an assignment using
measurements collected while its constituent plans run concurrently, because
their shared-resource interference cannot be inferred reliably from isolated
profiles alone. It first minimizes the stall of the stream that gates
application progress and then optimizes the remaining streams subject to their
declared constraints. TailGuard-GDS may therefore accept a modest slowdown in
a background checkpoint to protect a foreground parameter reload, but rejects
an assignment that would cause the checkpoint to miss its deadline or fall
below its required bandwidth.

Placement enters the policy because GDS and host-staged delivery traverse the
hardware topology differently. GDS performance depends on the PCIe route
between the target GPU and the NVMe devices holding the requested file blocks,
including any shared switches or CPU root complexes. Host-staged delivery
instead depends on both the storage-to-host-memory path and the
host-memory-to-GPU copy, consuming NUMA-local memory bandwidth and GPU
copy-engine resources. The distinction is especially important for striped
files whose blocks span NVMe devices with different locality relative to a GPU.
Unlike GPUDirect RDMA, where a communication runtime can often select a NIC or
rail close to the target GPU, a GDS request is constrained by the file system,
mount point, RAID layout, and existing file placement. TailGuard-GDS therefore
selects a placement only when the framework exposes alternatives, such as
eligible mounts, GPU groups, or CPU/NUMA affinity settings. When placement is
fixed, TailGuard-GDS uses it as context for selecting the backend and request
shape; it does not migrate files or repartition storage at runtime.

If no assignment satisfies all declared performance constraints, TailGuard-GDS
follows an ordered fallback procedure: (1) it retains the current verified plan
if that plan can still complete the required GPU-buffer operation, even when a
latency or throughput target cannot be met; (2) if the current plan is
unavailable or fails, it activates a conservative plan that has already been
validated for the phase; and (3) if neither plan can complete the operation
correctly, it returns the failure to the framework rather than masking it as
normal completion. When execution continues through either fallback path,
TailGuard-GDS reports the constrained state to the framework, including the
latency, throughput, placement, or backend-availability requirement that could
not be satisfied. The framework may use this signal to delay later checkpoints,
reduce background I/O, adjust prefetch or offload granularity, or exclude the
degraded observation from normal profile updates. How the framework reschedules
subsequent work using this signal is outside the scope of this paper.

\section{Implementation}

We implement TailGuard-GDS as a two-part runtime between AI frameworks and
GPU-storage backends. A Python control plane receives phase events from
framework adapters, constructs the context \(w\), consults the profile, applies
the selection and switching guards, and records phase completion. A native
C++/CUDA data plane provides a common GPU-buffer I/O interface for the GDS and
host-staged backends. Framework adapters invoke the runtime at existing storage
boundaries, while model code remains independent of backend selection. This
separation allows the same policy implementation to control different
frameworks and data-movement backends without changing their
application-visible completion semantics.

The native data plane exposes common \texttt{read-into} and
\texttt{write-from} operations over the same CUDA allocation. The GDS backend
opens the target file with direct I/O, registers the file handle and GPU buffer
with cuFile, and invokes \texttt{cuFileRead} or \texttt{cuFileWrite};
initialization succeeds only when compatibility fallback is disabled. The
host-staged backend accesses the same logical file range through direct
\texttt{pread} or \texttt{pwrite} operations on a reusable pinned host buffer.
A read is followed by an asynchronous host-to-device copy, whereas a write
first performs the device-to-host copy before issuing the file operation. Both
backends report completion only after the full byte range has reached the
required endpoint, and both apply the same optional durability operation to
writes. They therefore preserve a common GPU-buffer contract despite using
different internal data paths.

TailGuard-GDS amortizes setup costs by caching CUDA allocations, pinned staging
buffers, CUDA streams, cuFile registrations, and open file handles across
repeated phase invocations. A backend transition is serialized by the control
plane: it first drains outstanding requests from the active backend, registers
all live GPU endpoint buffers with the candidate backend, and activates the new
plan only after every registration succeeds. If registration fails,
TailGuard-GDS removes any partial candidate state and leaves the previous
backend active. An I/O failure after activation invokes the ordered fallback
procedure described in Section~\ref{sec:calibration-safety}. Consequently, an
application request observes either the previous verified plan or a fully
initialized replacement, rather than a partially completed transition.

We implement framework adapters for PyTorch FSDP and DeepSpeed ZeRO-3
\cite{fsdp2023,deepspeed2020,zero2020}. The
PyTorch adapter binds each framework-owned, rank-local FSDP shard to
TailGuard-GDS and invokes reload and offload operations at the corresponding
pre-forward and post-optimizer boundaries. The DeepSpeed adapter installs hooks
in the ZeRO-3 parameter coordinator and the parameter and optimizer NVMe
swappers, exposing their natural fetch, release, swap-in, and swap-out events to
the TailGuard control plane. Within the same Transformer training loop,
TailGuard-GDS applies phase- and shape-aware selection to GPU-buffer checkpoint
and offload operations executed through its data plane. Both adapters emit a
common phase record containing the context, selected plan, phase stall,
achieved throughput, guard outcome, and fallback state.

\section{Evaluation}
\label{sec:evaluation-methodology}

\textbf{Evaluation overview.}
We organize the evaluation around four questions.
\textbf{Q1: Path behavior and mechanism.} How and why do GDS direct and
host-staged delivery differ across request shapes, offered loads, and
GPU-storage placements?
\textbf{Q2: Policy effectiveness.} Can TailGuard-GDS use phase context and
measured tail behavior to select the backend, request shape, and available
placement that reduce application-visible stall while sustaining the required
data rate?
\textbf{Q3: Application impact.} Does this selection reduce storage-phase
stall and end-to-end step latency in multi-rank PyTorch workloads and
eight-GPU DeepSpeed ZeRO-3 training?
\textbf{Q4: Robustness and runtime cost.} How do TailGuard's decisions and
benefits change under multi-GPU contention, resource pressure, workload
variation, and topology changes, and what costs arise from calibration,
switching, and fallback?
We answer these questions progressively, moving from path-level measurements
and mechanism analysis to end-to-end application experiments.

\subsection{Experimental Setup}

\begin{figure}[t]
  \centering
  \includegraphics[width=\columnwidth]{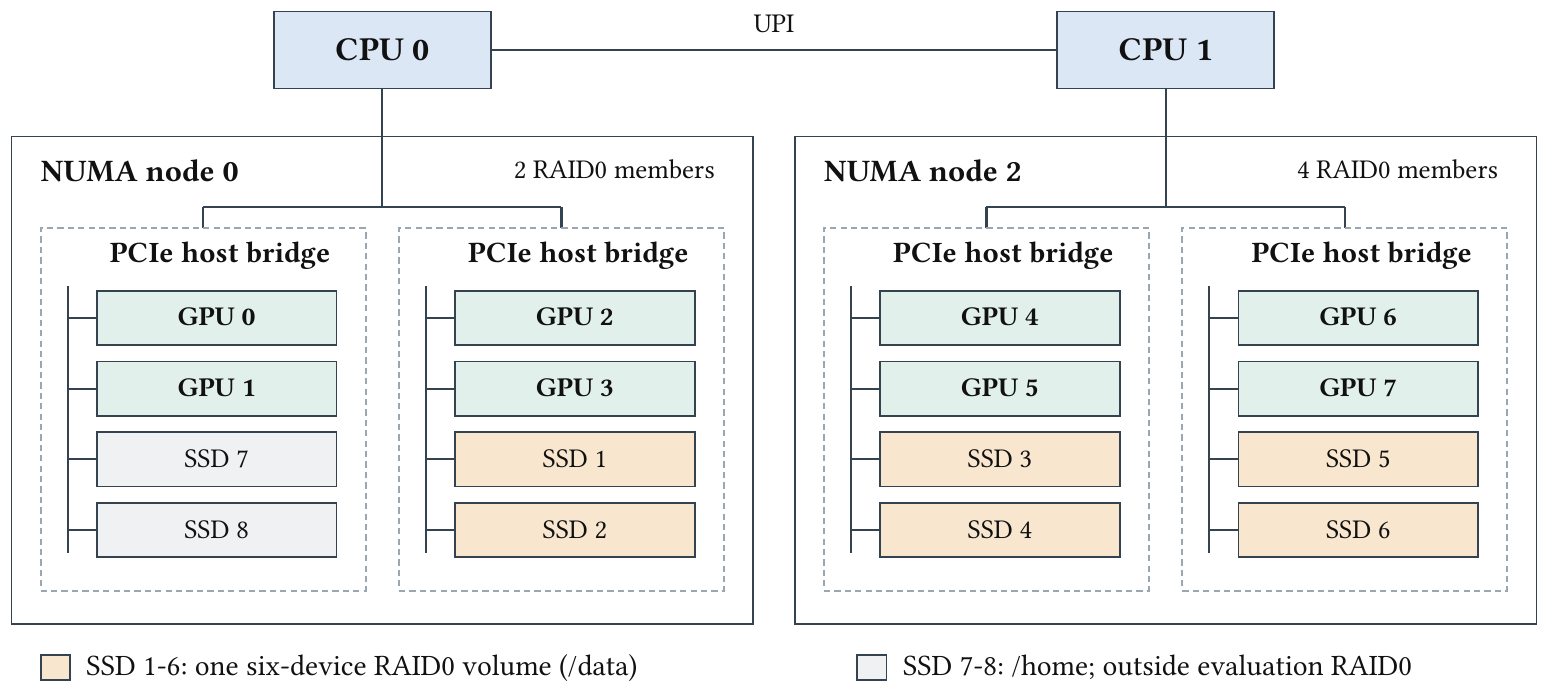}
  \caption{Evaluation topology. CPU~0 and CPU~1 are connected by UPI; only the
  NUMA domains hosting the evaluated GPUs and storage devices are shown.
  GPU0--GPU3 and two RAID0 members (SSD~1--SSD~2) attach to NUMA node~0,
  whereas GPU4--GPU7 and four RAID0 members (SSD~3--SSD~6) attach to NUMA
  node~2. SSD~7--SSD~8 host \texttt{/home} and are excluded from the
  evaluation RAID0. PCIe host-bridge connectivity is schematic.}
  \Description{Two-socket evaluation topology. GPU0 through GPU3 and two RAID
  members attach to NUMA node 0, while GPU4 through GPU7 and four RAID members
  attach to NUMA node 2; two additional SSDs host the home file system and are
  excluded from the evaluated RAID.}
  \label{fig:platform-topology}
\end{figure}

\textbf{Experimental platform.}
We evaluate TailGuard-GDS on a server with eight NVIDIA H100 80-GB GPUs, two
Intel Xeon Platinum 8558 processors, 96 physical cores, and four NUMA nodes.
GPU0--GPU3 are attached to NUMA node 0, while GPU4--GPU7 are attached to NUMA
node 2. The storage volume used in the experiments is an ext4 file system over
a six-device NVMe RAID0 array, with two RAID members attached to NUMA node 0
and four attached to NUMA node 2. Figure~\ref{fig:platform-topology} labels
these RAID members SSD~1--SSD~6: SSD~1--SSD~2 attach to NUMA node~0, while
SSD~3--SSD~6 attach to NUMA node~2. Two additional devices, SSD~7--SSD~8, host
\texttt{/home} and are outside the evaluation RAID0. This asymmetric topology
provides GPU-storage paths with different PCIe and NUMA locality. We evaluate
both single-GPU execution and concurrent execution across all eight GPUs to
separate individual-path behavior from shared-resource contention and scaling
effects. The server runs Linux 5.15, CUDA 13.0, NVIDIA driver 580.65.06, and
cuFile/GDS 1.15.

\textbf{GPU-delivery paths.}
All path comparisons use TailGuard-GDS's native data plane and the two backends
described in Section~4. GDS direct issues \texttt{cuFileRead} and
\texttt{cuFileWrite} on registered GPU memory, while host staging combines
direct file I/O on reusable pinned memory with asynchronous CUDA copies.
Request size, queue depth, batching, and concurrency are configured
independently for each candidate plan. Within every comparison, the two paths
operate on identical files, offsets, payloads, GPU buffers, transfer volumes,
and write-durability requirements. Read timing ends when the requested data is
available to dependent GPU work; write timing ends when the data has reached
storage and the configured durability operation has completed.

\textbf{GDS configuration.}
All GDS measurements use cuFile/GDS 1.15 with \texttt{nvidia\_fs} 2.29.4 loaded
and \texttt{allow\_compat\_\allowbreak mode=false}. Each run begins after successful
file-handle registration, GPU-buffer registration, and a direct-I/O preflight
on the selected GPU and storage path. The reported GDS results therefore
represent native transfers between NVMe storage and GPU memory.

\textbf{E1: Microbenchmark path characterization.}
To answer Q1, we first measure a saturation matrix containing all 36
combinations of operation (read or write), access pattern (sequential or
random), request size (4~KiB, 64~KiB, or 1~MiB), and worker count (1, 4, or 8).
Each combination operates on a 4~GiB direct-I/O file striped across the
six-device RAID0 array and is repeated three times for both delivery paths.
These runs measure the attainable bandwidth, IOPS, and latency distribution of
each path. We then conduct matched-load experiments. For each request shape, we
measure the sustainable capacities \(C_{\mathrm{gds}}\) and
\(C_{\mathrm{host}}\), define
\(C_{\min}=\min(C_{\mathrm{gds}},C_{\mathrm{host}})\), and pace both paths at
identical offered rates of \(0.5C_{\min}\), \(0.75C_{\min}\), and
\(0.9C_{\min}\). Comparing the paths at the same offered rate separates
differences in their latency distributions from queueing caused by unequal
request rates.

\textbf{E2: Multi-rank PyTorch path selection.}
To answer Q2 and connect path selection to Q3, we run a four-rank PyTorch FSDP
workload \cite{fsdp2023} in which every rank executes forward propagation, backward
propagation, and an optimizer update on an FSDP-sharded model. TailGuard-GDS
reloads the framework-owned rank-local shard before dependent computation and
offloads it after the optimizer phase. All ranks synchronize at the end of each
step, so a storage delay on the slowest rank becomes visible in end-to-end step
latency. We vary shard size, issued request shape, GPU-storage placement, and
the number of participating ranks while keeping the model and compute schedule
unchanged. The experiment measures whether phase- and shape-aware plan
selection reduces reload and offload stall, GPU idle time, and synchronized
step latency.

\textbf{E3: End-to-end DeepSpeed ZeRO-3 training.}
To answer Q3, we run eight-GPU DeepSpeed ZeRO-3 training with Hugging Face
implementations of BERT, GPT-2, OPT, LLaMA, and T5
\cite{deepspeed2020,zero2020}. Each workload executes
forward and backward propagation, optimizer updates, ZeRO partitioning,
inter-rank communication, parameter and optimizer NVMe activity, and GPU-buffer
checkpoint and offload phases within the same training loop. The DeepSpeed
adapter records the natural ZeRO-3 storage events, while TailGuard-GDS selects
the backend and issued request shape for eligible GPU-buffer storage phases.
For each model architecture, we measure 256 training steps and compare
TailGuard-GDS with fixed GDS, fixed host staging, and a platform-tuned DeepSpeed
AIO configuration. We report storage-phase stall, GPU idle time, synchronized
step latency, and training throughput.

\textbf{E4: Robustness and runtime cost.}
To answer Q4, we vary four dimensions of the execution environment: (1) the
number of concurrently active GPUs, (2) request shape and offload concurrency,
(3) GPU-storage placement, and (4) competing CPU, host-memory, and CUDA-copy
pressure. We introduce resource-pressure transitions within a running workload
to determine when an existing profile becomes stale and whether recalibration
selects a better plan. We also measure profile lookup, calibration, and
backend-switching costs, together with cumulative decision regret over extended
execution. Finally, checksum-preserving path switches and injected registration
and I/O failures evaluate completion correctness and fallback behavior. These
experiments quantify both the adaptability and the runtime cost of
TailGuard-GDS across model architectures and workload intensities.

\textbf{Metrics and statistical treatment.}
At the path level, we report achieved bandwidth or IOPS together with p50, p95,
p99, and p999 per-request latency. We use p999 as the primary path-level tail
metric because an extreme I/O delay can become a rank-level straggler. For each
microbenchmark configuration, latency quantiles are computed from post-warmup
I/O samples, and repeated runs are summarized by their median. At the
application level, we report storage-phase stall, GPU idle time, synchronized
step p95, and training throughput; unless stated otherwise, step completion is
determined by the slowest participating rank. The resource-pressure transition
experiment instead retains rank-local step durations: each plotted p95 pools 32
measured steps from each of eight ranks (256 samples), rather than first reducing
each logical step to its slowest rank. Primary application configurations contain
at least three complete runs. Their 95\% confidence intervals are computed with a hierarchical
nonparametric bootstrap that first resamples runs and then resamples post-warmup
steps within each selected run. Tail metrics are reported together with
completed work and throughput, and every selected plan must satisfy the phase's
\(B_{\min}(w)\) requirement.

\textbf{Baselines and ablations.}
Fixed GDS and fixed host staging keep the transfer backend constant throughout
execution. We evaluate both backends over the same request-shape configurations
available to TailGuard and report their strongest fixed configurations. Tuned
DeepSpeed AIO uses the framework's native asynchronous NVMe implementation,
with block size, queue depth, and worker count selected through a platform-wide
parameter sweep. Best-static is the fixed end-to-end policy with the lowest
measured application tail over the evaluated workload. An offline phase oracle
uses the complete measurement results to select the lowest-stall candidate for
each application phase, representing the performance available with perfect
knowledge of phase behavior. To isolate TailGuard's design choices, phase-only
assigns one fixed plan to each application phase, backend-only changes the
transfer backend while holding request shape fixed, and shape-only changes
request shape while holding the backend fixed. TailGuard jointly selects
backend and request shape and incorporates placement when the framework exposes
multiple eligible alternatives.

\textbf{Experimental controls.}
Comparisons within each experiment use the same model and training
configuration, input sequence, logical file ranges, transfer volume, file
layout, rank-to-GPU mapping, and available placement choices. Measurement
begins after one-time framework initialization, resource creation, and backend
warmup. Recurring runtime costs, including plan lookup, calibration,
recalibration, and path switching, remain inside the measured interval. We
rotate the execution order of competing policies across repeated runs and
synchronize all ranks before entering each timed region.

\subsection{E1: Path Performance Is Workload Dependent}

Figure~\ref{fig:semantic-equiv} answers Q1 under saturated load. Host staging
minimizes p999 latency for most request shapes, whereas GDS becomes preferable
for selected 64~KiB write configurations. More importantly, changing the
operation type, request size, or concurrency can alter the relative ordering of
the two paths. The marked cases also show that the path with higher bandwidth
can have worse tail latency. Requiring every selected plan to preserve at least
90\% of the best achievable bandwidth does not change the lower-p999 choice for
any evaluated request shape, indicating that the tail improvement does not
require a substantial throughput sacrifice. A fixed GDS or fixed host-staging
policy can increase p999 by up to \(6.92\times\) and \(4.20\times\),
respectively. These results show that GPU-storage path selection must account
for request shape and the phase's throughput requirement rather than relying on
backend identity or bandwidth alone.

Figure~\ref{fig:matched-load} repeats the comparison at identical target IOPS
chosen from the throughput range that both paths can sustain. The median
absolute target error is 0.299\%. In 100 of 108 cases, the GDS-to-host
achieved-IOPS ratio lies within \(0.98\times\)--\(1.02\times\); the remaining
eight cases occur under high offered load, where one path does not sustain the
common target and the observed ratio spans \(0.883\times\)--\(1.041\times\).
Among the 100 achieved-load-matched cases, host staging has lower p999 in 84
cases. The GDS-to-host p999 ratio reaches \(3.60\times\) within the experiment.
The path-dependent tail difference therefore persists when GDS and host
staging process comparable achieved loads; unequal saturation throughput alone
does not explain the result.

\begin{figure}[t]
  \centering
  \includegraphics[width=\columnwidth]{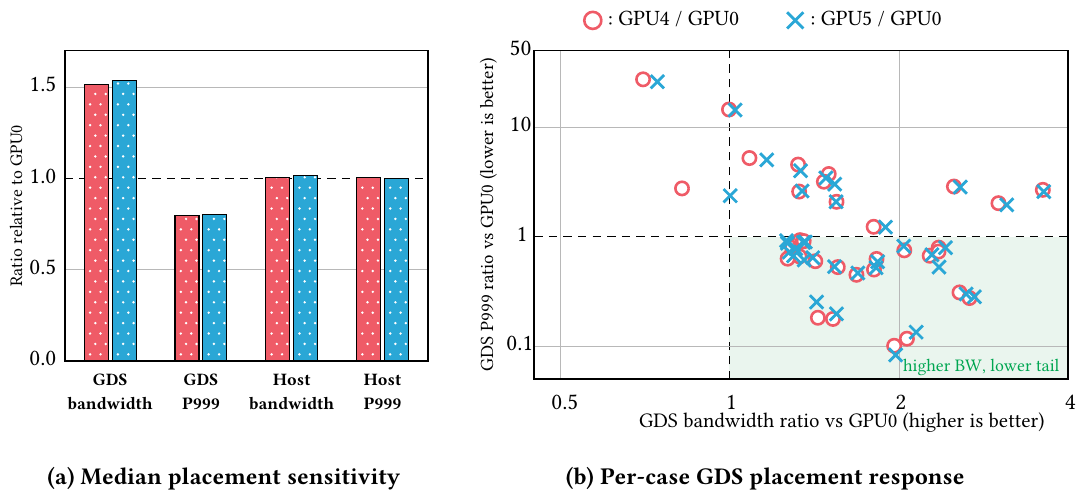}
  \caption{Placement affects GDS selectively. Bars show paired medians relative
  to GPU0; points retain all operation, size, and concurrency cases. Lower p999
  and higher bandwidth are better.}
  \Description{Placement sensitivity results. GDS bandwidth generally rises
  and GDS p999 generally falls on NUMA-2 GPUs relative to GPU0, while the
  host-staged path remains close to its GPU0 result.}
  \label{fig:gpu-locality}
\end{figure}

Figure~\ref{fig:gpu-locality} shows that placement affects the paths
differently. Running the same workloads on the NUMA-2 GPUs, which share a NUMA
domain with four of the six RAID members, increases median GDS bandwidth by
\(1.51\text{--}1.54\times\) and reduces median GDS p999 to approximately
\(0.80\times\) of the GPU0 result. In contrast, host-staged bandwidth and p999
remain within 1.5\% of their GPU0 values. The case-level results in panel (b)
show that roughly two thirds of the request shapes improve both GDS bandwidth
and tail latency, while others gain bandwidth without reducing the tail. Under
matched load, the fraction of cases in which GDS provides lower p999 increases
from 16\% on GPU0 to 56\% on GPU4, and the preferred path changes in 61\% of
the comparisons. Placement therefore changes both the absolute performance of
GDS and the relative ordering of the two backends, motivating its inclusion in
TailGuard's decision context.

\subsection{E2: From Path Diversity to Application Gains}

\begin{figure}[t]
  \centering
  \includegraphics[width=\columnwidth]{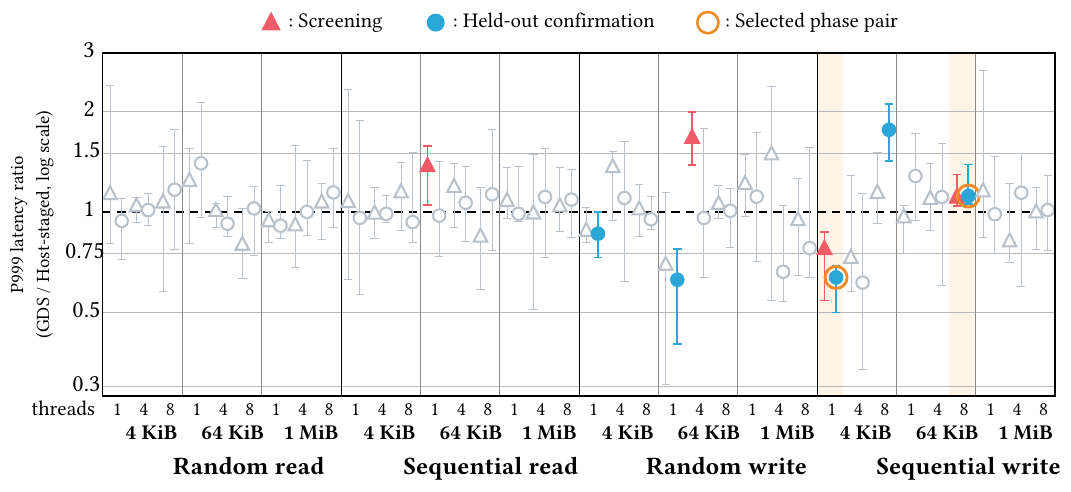}
  \caption{Candidate screening and held-out confirmation. Triangles summarize
  repeats 1--10 and circles summarize disjoint repeats 11--20. Each point is
  the median GDS/host-staged p999 ratio across ten repeats; error bars are
  95\% bootstrap intervals of that median from 20,000 resamples. Filled
  markers pass the consistency, confidence, execution-order, and load guards;
  orange outlines identify the opposite-winner pair selected before adaptive
  evaluation.}
  \Description{Screening and held-out confirmation ratios across request
  shapes. Most candidates are near parity or unstable, while two sequential
  write shapes retain opposite and statistically confirmed backend
  preferences.}
  \label{fig:native-candidate-gate}
\end{figure}

Figure~\ref{fig:native-candidate-gate} examines whether performance differences
identified during calibration persist in an independent measurement set. Each
marker reports the GDS-to-host p999 ratio: values below one favor GDS, whereas
values above one favor host staging. We perform ten screening repetitions
followed by ten disjoint confirmation repetitions and retain candidates that
satisfy the consistency, confidence, execution-order, and load criteria
described in Section~\ref{sec:design}. Most request shapes remain close to
parity or change ordering between the two measurement sets. Two request shapes
retain stable but different backend preferences. A 4~KiB sequential write with
one thread achieves a confirmation ratio of \(0.636\times\) and favors GDS in
every confirmation run, whereas a 64~KiB sequential write with eight threads
achieves a ratio of \(1.116\times\) and favors host staging in nine of ten runs.
These results establish that backend preference can remain stable yet differ
across request shapes. They are calibration examples rather than the phase map
used by the following FSDP experiment. For that experiment, the launcher fixes
and the trace verifies an application-specific map: the pre-forward parameter
reload is a read delivered through host staging, and the post-optimizer
parameter offload is a write delivered through GDS. Each phase issues one
request equal to the rank-local shard size. Appendix~\ref{app:phase-maps}
lists the complete mapping and its source records.

\begin{figure}[t]
  \centering
  \includegraphics[width=\columnwidth]{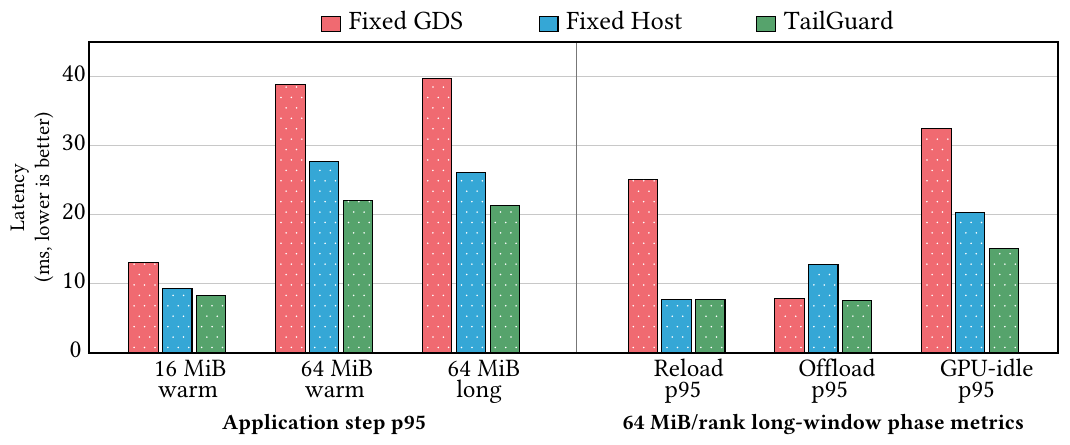}
  \caption{Four-rank PyTorch FSDP application results. The left group reports
  step p95 across shard sizes and measurement windows; the right group
  separately reports reload, offload, and GPU-idle p95 for the
  64~MiB/rank long-window run. Warm runs contain 6--8 measured steps per rank
  and repeat; the long run contains 32. These percentiles are computed
  independently and are not additive. Lower is better.}
  \Description{Grouped bars comparing Fixed GDS, Fixed Host, and TailGuard for
  four-rank FSDP. TailGuard has the lowest application step p95 in all three
  runs and lowers the long-window GPU-idle p95 while using different paths for
  reload and offload.}
  \label{fig:fsdp-application}
\end{figure}

Figure~\ref{fig:fsdp-application} evaluates whether this application-specific
phase map reduces the latency of synchronized PyTorch FSDP training steps.
TailGuard assigns host staging to parameter reload and GDS to parameter
offload, whereas each fixed baseline applies one backend to both phases. The
left group shows that this phase-specific assignment reduces step p95 by 10.7\%
at 16~MiB per rank and 19.6\% at 64~MiB per rank relative to the strongest
fixed-backend policy. The 64~MiB long-window run repeats the experiment over a
longer measurement window and obtains an 18.4\% reduction in step p95, with a 95\%
confidence interval of 18.3--18.6\%. TailGuard also reduces mean step latency
by 19.0\% and GPU-idle p95 by 25.3\%. The corresponding reductions in
storage-phase stall, GPU waiting time, and synchronized step latency show how
path-level decisions translate into application-level improvement.

\subsection{E3: End-to-End DeepSpeed ZeRO-3 Training}

\begin{table*}[t]
\centering
\caption{Repeated 8GPU, 256-step DeepSpeed ZeRO-3 runs. Cells are mean step p95
in ms (\(\pm\) standard deviation); gains are mean reductions. Lower is better.}
\label{tab:app-longwindow-models}
\scriptsize
\setlength{\tabcolsep}{5pt}
\begin{tabular}{lrrrrrr}
\toprule
Model & Fixed GDS & Fixed host & Phase-only & TailGuard & Gain vs host & Gain vs phase \\
\midrule
BERT-MLM & $863.1{\pm}5.7$ & $591.5{\pm}46.8$ & $852.8{\pm}1.9$ & $538.1{\pm}16.6$ & 8.8\% & 36.9\% \\
GPT-2 & $802.7{\pm}1.0$ & $571.0{\pm}73.4$ & $798.6{\pm}8.2$ & $493.7{\pm}7.1$ & 12.7\% & 38.2\% \\
OPT & $832.5{\pm}1.9$ & $624.3{\pm}15.2$ & $816.8{\pm}1.4$ & $539.8{\pm}18.8$ & 13.6\% & 33.9\% \\
LLaMA & $804.0{\pm}4.7$ & $611.2{\pm}33.0$ & $801.0{\pm}10.0$ & $461.8{\pm}54.1$ & 24.6\% & 42.4\% \\
T5 & $1025.0{\pm}3.3$ & $802.4{\pm}76.7$ & $1014.0{\pm}6.1$ & $733.6{\pm}55.5$ & 7.9\% & 27.7\% \\
\bottomrule
\end{tabular}
\end{table*}

Table~\ref{tab:app-longwindow-models} reports repeated 8GPU, 256-step DeepSpeed
ZeRO-3 runs across five Transformer architectures. TailGuard achieves the
lowest mean step p95 for
every model. Fixed host staging consistently outperforms fixed GDS and
phase-only selection, confirming that the transfer backend materially affects
the training critical path. TailGuard further reduces step p95 by
7.9--24.6\% relative to fixed host staging, showing that selecting one globally
stronger backend cannot capture the request-shape differences between training
phases. Relative to fixed GDS and phase-only selection, the reductions reach
28.4--42.6\% and 27.7--42.4\%, respectively. This trend holds across the
encoder-only BERT workload, the decoder-only GPT-2, OPT, and LLaMA workloads,
and the encoder-decoder T5 workload.

\begin{figure}[t]
  \centering
  \includegraphics[width=\columnwidth]{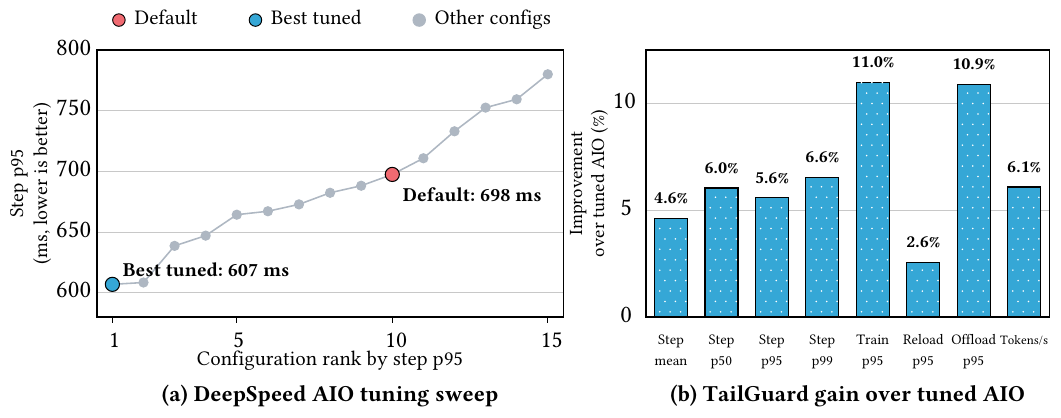}
  \caption{DeepSpeed AIO baseline tuning and paired TailGuard comparison.
  Panel (a) ranks 15 configurations using 512 rank-step samples per
  configuration. Panel (b) is an independent run with 1,024 rank-step samples
  per policy and uses the selected AIO parameters for both policies.
  \emph{Train p95} is the p95 duration of model forward, backward, optimizer
  step, and the following GPU synchronization, excluding the separately timed
  reload and offload phases. Latency gains are reductions; the tokens/s gain
  is an increase.}
  \Description{Two-panel DeepSpeed AIO evaluation. The first panel ranks 15 AIO
  configurations by step p95, and the second shows that TailGuard lowers step
  and training latency while increasing estimated throughput relative to the
  selected AIO configuration.}
  \label{fig:aio-tuning}
\end{figure}

Figure~\ref{fig:aio-tuning} compares TailGuard with a tuned DeepSpeed AIO
baseline. Panel (a) evaluates 15 AIO configurations spanning block size, queue
depth, worker count, submission mode, and event overlap. The best configuration
reduces step p95 from 697.5 to 607.0~ms, a 13.0\% improvement over the default
configuration. Panel (b) performs an independent paired run using the selected
AIO parameters for both the native baseline and TailGuard's host-staged
candidate. Within this run, TailGuard reduces step p95 from 668.9 to 631.5~ms
and offload p95 from 369.8 to 329.5~ms, corresponding to reductions of 5.6\%
and 10.9\%, while increasing estimated throughput by 6.1\%. These results show
that runtime backend and request-shape selection provides benefits beyond
static tuning of the framework's native asynchronous I/O path.

\begin{table*}[t]
\centering
\caption{Strategy ablation on live 8GPU DeepSpeed runs. Cells are extra step-p95
cost relative to joint backend-and-shape selection; lower is better.}
\label{tab:strategy-ablation}
\scriptsize
\setlength{\tabcolsep}{5pt}
\begin{tabular}{lrrrrrrr}
\toprule
Model & Fixed GDS & Fixed host & Phase-only & Backend-only & Shape-only & Host-only shaped & GDS-only shaped \\
\midrule
BERT-MLM & 55.6\% & -3.3\% & 52.1\% & 5.9\% & 52.4\% & 0.8\% & 57.8\% \\
GPT-2 & 67.8\% & 24.6\% & 65.3\% & 11.9\% & 71.4\% & 17.0\% & 69.8\% \\
\bottomrule
\end{tabular}
\end{table*}

Table~\ref{tab:strategy-ablation} reports the strategy ablation results.
Relative to joint backend-and-shape selection, phase-only increases step p95 by
52.1--65.3\%, while shape-only increases it by 52.4--71.4\%. GDS-only shaping
produces similarly large overheads of 57.8--69.8\%, indicating that request
shaping cannot compensate for an unfavorable backend. Backend-only selection
substantially narrows the gap but remains 5.9--11.9\% slower because it cannot
adapt the transfer shape of each phase. Host-only shaping approaches joint
selection on BERT, with a 0.8\% difference, but remains 17.0\% slower on GPT-2.
These results show that backend selection contributes the largest performance
gain, while joint request-shape selection is necessary to sustain that gain
across different application workloads.

\subsection{E4: Robustness Across Operating Regimes}

\begin{figure}[t]
  \centering
  \includegraphics[width=\columnwidth]{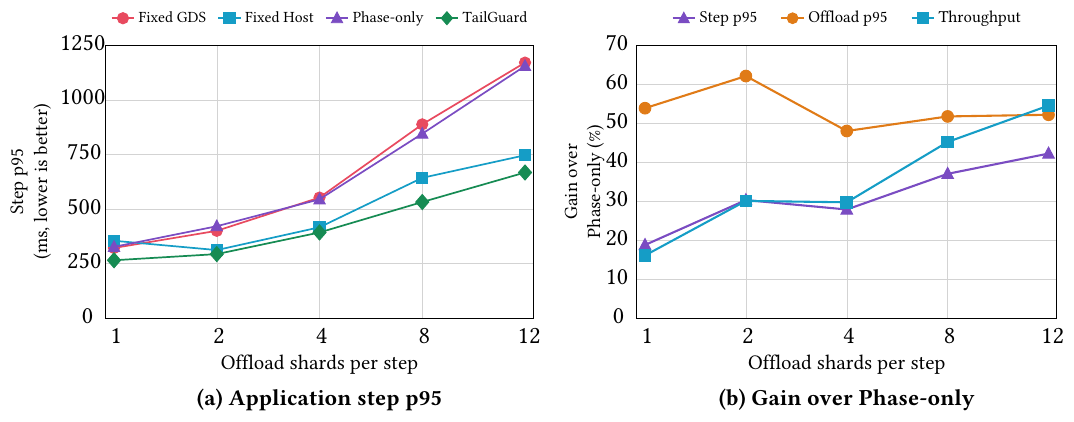}
  \caption{Eight-GPU BERT offload-pressure sweep. The shard counts are
  categorical workload settings. More shards place more persistent I/O on the
  step critical path. In panel (b), step and offload gains are latency
  reductions, whereas throughput gain is an increase; all use Phase-only as
  the baseline.}
  \Description{BERT pressure sweep over increasing offload shard counts.
  TailGuard's step p95 grows more slowly than the fixed and phase-only policies,
  and its relative latency and throughput gains increase as offload pressure
  rises.}
  \label{fig:pressure-sweep8}
\end{figure}

Figure~\ref{fig:pressure-sweep8} answers Q4 by increasing the number of shards
offloaded to NVMe in each training step from one to twelve. As the persistent-I/O
pressure grows, the step p95 of fixed GDS and phase-only selection increases
most rapidly, reaching approximately 1.15--1.17~s at twelve shards. At the same
pressure, TailGuard achieves 670~ms, compared with 730~ms for fixed host
staging. Panel (b) shows that TailGuard's step-p95 reduction over Phase-only increases
from 18.8\% to 42.2\%, while its reduction over fixed host staging ranges from
5.7\% to 25.2\%. GPT-2 exhibits the same trend across one, eight, and twelve
shards, with TailGuard reducing step p95 by 24.3\% relative to fixed host
staging at the highest pressure. These results show that the benefit of joint
path and request-shape selection increases as storage activity occupies a
larger fraction of the training critical path.

\begin{figure*}[t]
  \centering
  \includegraphics[width=0.98\textwidth]{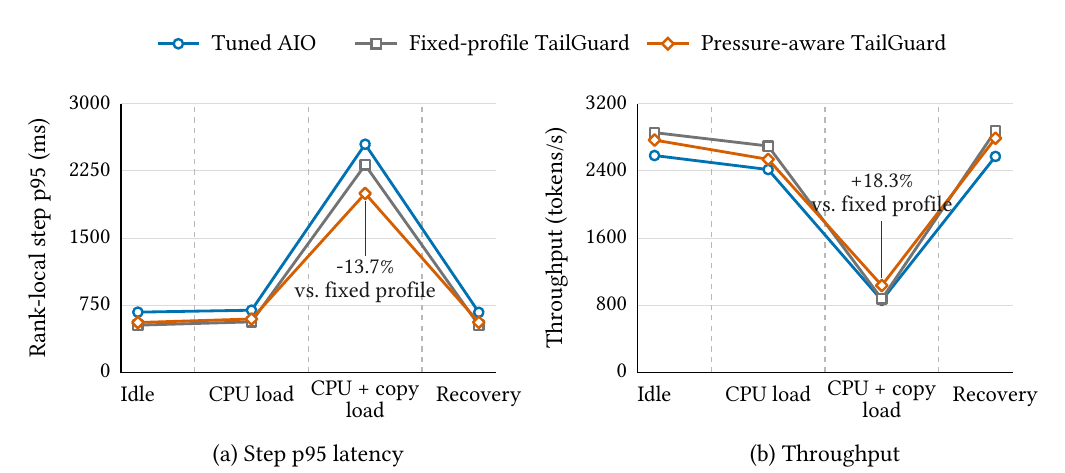}
  \caption{Performance across four consecutive resource-pressure windows in
  three separate eight-GPU runs. Each policy follows the same sequence: idle,
  CPU load, CPU plus CUDA-copy load, and recovery. Fixed-profile TailGuard
  retains the request shape calibrated in the initial idle window, whereas
  Pressure-aware TailGuard uses the pressure-state label supplied at each window
  boundary to select the corresponding request shape. Each point summarizes one
  resource-pressure window. Step p95 is computed over 256 pooled rank-local
  durations (32 measured steps per rank); connecting lines indicate window order
  rather than continuous-time measurements. Annotations compare Pressure-aware TailGuard with
  Fixed-profile TailGuard under combined pressure.}
  \Description{Step p95 and throughput across idle, CPU pressure, combined CPU
  and CUDA-copy pressure, and recovery windows. Pressure-aware TailGuard
  outperforms the fixed profile under combined pressure and returns toward its
  idle performance during recovery.}
  \label{fig:pressure-aware8}
\end{figure*}

Figure~\ref{fig:pressure-aware8} evaluates pressure-conditioned profile selection as each run moves from
idle to CPU load, then to combined CPU and CUDA-copy load, and finally back to
idle. At each boundary, the experiment supplies the pressure-state label,
selects the corresponding profile, activates the requested pressure, waits five
seconds, and synchronizes the ranks before starting the timed window. The figure
therefore reports performance within each post-transition measurement window;
it does not include automatic pressure-detection or adaptation time. Under
combined pressure, retaining the idle profile raises step p95 to 2.32~s and
limits throughput to 876 tokens/s. Pressure-aware TailGuard
changes the active request size and concurrency, lowering step p95 to 2.00~s,
13.7\% below the fixed profile, while increasing throughput by 18.3\%. The
backend remains host-staged and placement is unchanged in this experiment, so
the difference isolates pressure-conditioned request-shape selection. Relative to tuned
DeepSpeed AIO, Pressure-aware TailGuard reduces step p95 in all four windows and
provides its largest reduction, 21.5\%, under combined pressure. When
contention is removed, it restores the idle request shape and recovers both
latency and throughput. The transition experiment confirms that resource state
belongs in the decision context and that a stale profile should be revisited
when that state changes.

\begin{figure}[t]
  \centering
  \includegraphics[width=\columnwidth]{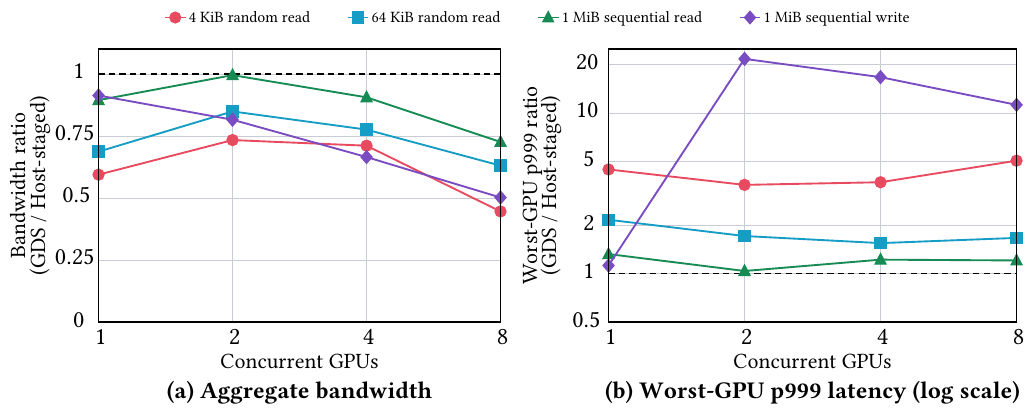}
  \caption{Strict pure-I/O scaling under one to eight concurrent GPUs. Ratios
  are GDS divided by host-staged delivery; the dashed line denotes parity.}
  \Description{GDS-to-host ratios as GPU concurrency increases from one to
  eight. Aggregate bandwidth remains at or below parity, while worst-GPU p999
  stays above parity and varies sharply with request shape and GPU count.}
  \label{fig:multigpu-contention}
\end{figure}

Figure~\ref{fig:multigpu-contention} examines how the relative performance of
GDS and host staging changes as the number of concurrent GPUs increases from
one to eight. Panel (a) shows that the GDS-to-host aggregate-bandwidth ratio
remains between \(0.45\times\) and \(1.00\times\), with host staging matching
or exceeding GDS across the evaluated configurations. Panel (b) reports the
worst-GPU p999 because synchronized multi-rank execution progresses at the pace
of its slowest rank. The GDS-to-host p999 ratio remains above one but varies
sharply with request shape and GPU concurrency. For 1~MiB sequential writes,
it increases from \(1.12\times\) with one GPU to \(21.7\times\) with two GPUs,
before reaching \(16.7\times\) and \(11.2\times\) with four and eight GPUs.
These results show that multi-GPU concurrency can amplify the tail-latency
difference between the two paths even when their aggregate-bandwidth ratio
changes modestly. TailGuard therefore includes the number and identity of
active GPUs, together with per-GPU concurrency, in its decision context.

\begin{figure*}[t]
  \centering
  \includegraphics[width=0.98\textwidth]{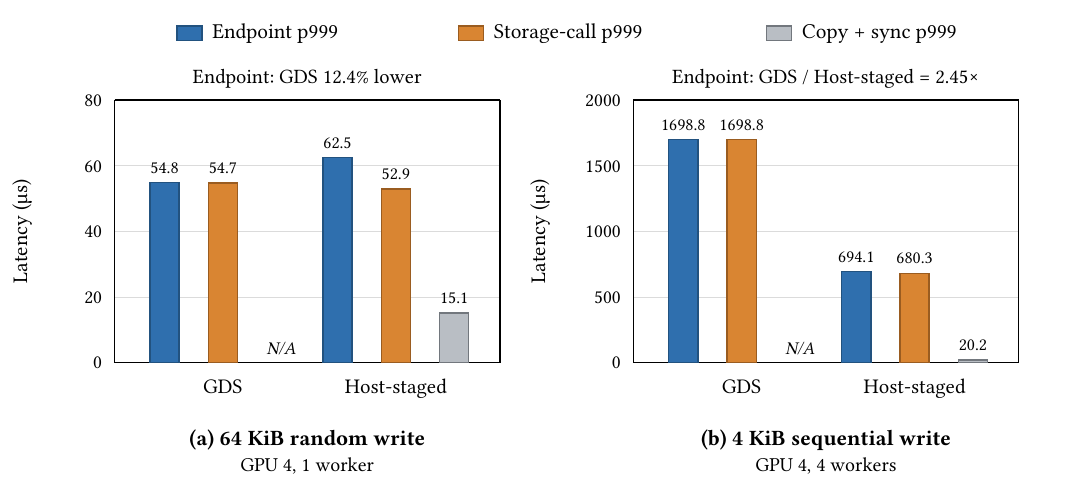}
  \caption{Independent endpoint and stage p999 values for (a) a 64~KiB random
  write on GPU~4 with one worker and (b) a 4~KiB sequential write on GPU~4
  with four workers. Endpoint timing begins immediately before path dispatch
  and ends after the selected path returns. Storage-call timing encloses
  synchronous \texttt{cuFileWrite} for GDS and \texttt{pwrite} for host
  staging; host copy-plus-synchronization timing encloses the device-to-host
  CUDA copy and stream synchronization before \texttt{pwrite}. Each value is
  the median of five run-level p999 estimates, with 12,000 requests per run in
  (a) and 20,000 in (b). Endpoint and stage p999 values are computed
  independently and are not additive. The panels use different y-axis scales.
  GDS has no copy stage, marked N/A. Annotations compare endpoint p999,
  reporting the GDS reduction relative to host staging in (a) and the
  GDS-to-host ratio in (b); the 12.4\% reduction is computed from the
  unrounded endpoint values 54.771 and 62.492~$\mu$s.}
  \Description{Endpoint, storage-call, and host copy-plus-synchronization p999
  for two request shapes. GDS has lower endpoint p999 for the random-write
  shape, while host staging has substantially lower endpoint p999 for the
  sequential-write shape; stage percentiles are shown independently.}
  \label{fig:component-decomposition}
\end{figure*}

Figure~\ref{fig:component-decomposition} compares the user-space timing
intervals for the two representative shapes. For the 64~KiB random write,
endpoint p999 is 54.771~$\mu$s for GDS and 62.492~$\mu$s for host staging, a
12.4\% reduction. Their storage-call p999 values are 54.707 and
52.905~$\mu$s, respectively, while the host copy-plus-synchronization p999 is
15.116~$\mu$s. For the 4~KiB sequential write, endpoint p999 is
1698.830~$\mu$s for GDS and 694.079~$\mu$s for host staging, a
$2.45\times$ GDS-to-host ratio. The corresponding storage-call p999 values are
1698.762 and 680.268~$\mu$s, while host copy plus synchronization remains
20.226~$\mu$s. Under this marginal-quantile definition, the stage p999 values
are independently computed summaries rather than components of one endpoint
tail request, and therefore do not sum to endpoint p999. The explicit
host-to-GPU copy alone does not explain the endpoint tail-latency ordering; in
the host-favored shape, the measured difference is primarily associated with
the storage-call interval.

\begin{figure}[t]
  \centering
  \includegraphics[width=\columnwidth]{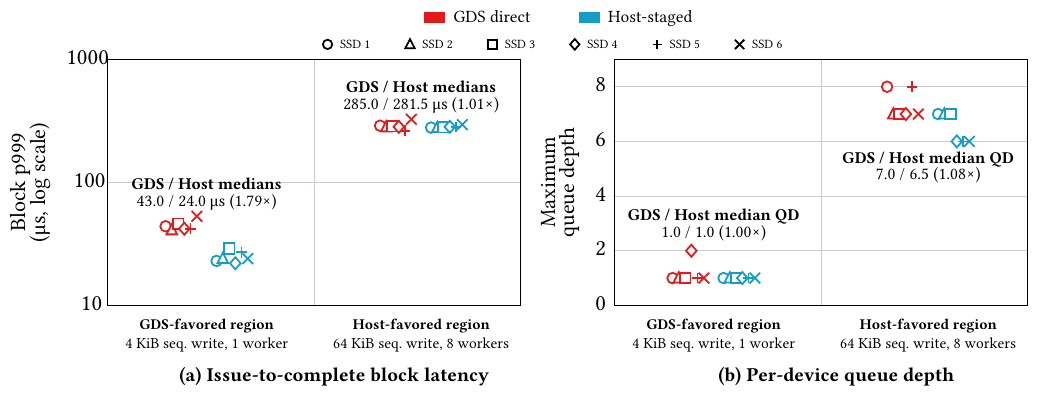}
  \caption{Block-layer diagnosis of one GDS-favored and one host-favored
  region. Color identifies the GPU-delivery path and marker shape identifies
  the six physical RAID members. Panel (a) reports issue-to-complete block
  p999; panel (b) reports maximum per-device queue depth. The common targets
  are 16.0~kIOPS for the 4~KiB sequential-write/one-worker region and
  7.32~kIOPS for the 64~KiB sequential-write/eight-worker region.}
  \Description{Per-SSD block-layer p999 and maximum queue depth for a
  GDS-favored region and a host-favored region. The block-layer ordering does
  not consistently match the endpoint tail-latency winner.}
  \label{fig:root-block}
\end{figure}

Figure~\ref{fig:root-block} continues the diagnosis below the user-space calls
by comparing block-layer behavior in one GDS-favored region and one
host-favored region. If NVMe service time or device queueing directly
determined the GPU-visible result, the path with lower endpoint p999 would also
exhibit lower block issue-to-complete latency or queue depth. In the
GDS-favored region, however, GDS achieves lower GPU-visible p999 despite having
\(1.79\times\) the median block p999 of host staging, while the two paths reach
the same median maximum queue depth. In the host-favored region, their median
block p999 and maximum queue depth are nearly identical, with ratios of
\(1.01\times\) and \(1.08\times\), respectively. The measured block service
and queueing metrics therefore do not track the endpoint ordering in these
representative regions. Together, Figures~\ref{fig:component-decomposition}
and~\ref{fig:root-block} show that neither the explicit host copy nor the
observed block-layer metrics alone explain the endpoint ordering, motivating
TailGuard's use of end-to-end GPU-visible tail latency as its selection signal.

\begin{figure}[t]
  \centering
  \includegraphics[width=\columnwidth]{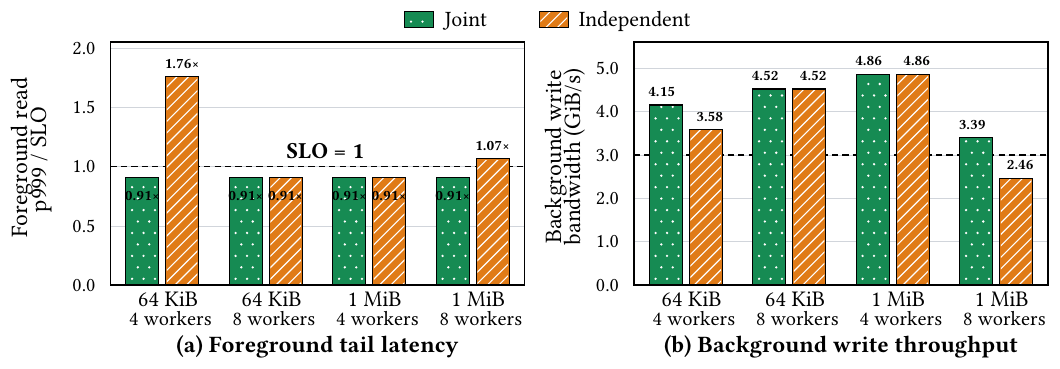}
  \caption{Joint versus independent path selection for a latency-critical
  foreground read and a bandwidth-constrained background write. Panel (a)
  reports foreground p999 normalized to its SLO; panel (b) reports background
  write bandwidth. Dashed lines denote the foreground SLO and the
  3.0~GiB/s write-bandwidth floor, respectively.}
  \Description{Foreground normalized p999 and background write bandwidth for
  joint and independent plan selection. Joint selection meets both the
  foreground latency SLO and background bandwidth floor across all request
  shapes, while independent selection violates one or both constraints.}
  \label{fig:mixed-stream-joint}
\end{figure}

Figure~\ref{fig:mixed-stream-joint} evaluates a latency-critical foreground
read running concurrently with a background write that must sustain at least
3.0~GiB/s. The independent baseline selects the best plan for each stream from
its isolated measurements, whereas TailGuard evaluates the two plans as a
joint assignment. TailGuard satisfies both the foreground p999 SLO and the
background bandwidth floor for all four foreground request shapes. Independent
selection, however, raises foreground p999 to \(1.76\times\) the SLO for the
64~KiB/four-thread configuration and to \(1.07\times\) the SLO for the
1~MiB/eight-thread configuration. In the latter configuration, it also
provides only 2.46~GiB/s of background bandwidth, below the required
3.0~GiB/s. These results show that plans selected from isolated stream
performance do not necessarily compose under shared storage contention.
TailGuard therefore selects the joint assignment that minimizes foreground
tail latency while preserving the background stream's bandwidth requirement.

\begin{figure}[t]
  \centering
  \includegraphics[width=\columnwidth]{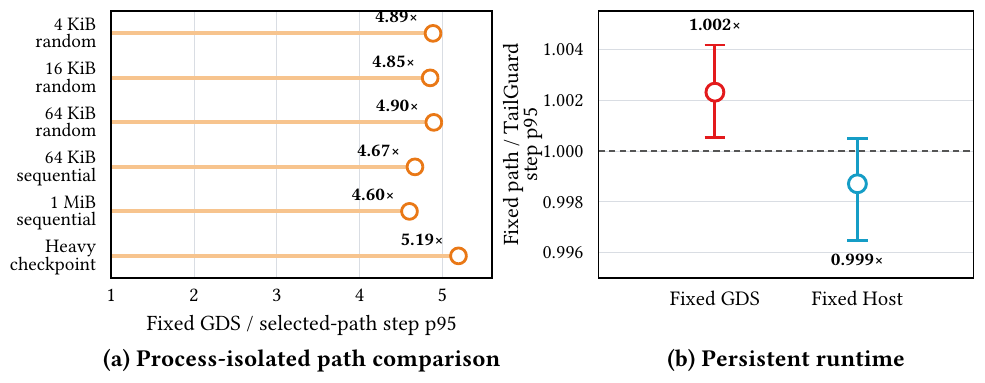}
  \caption{Runtime path comparison and steady-state control cost. Panel (a)
  reports Fixed GDS / selected-path step p95 for six workloads executed by the
  process-isolated runner. Panel (b) keeps the process, registrations, and GPU
  buffers alive and reports fixed-path/TailGuard step p95. Its 95\% confidence
  intervals bootstrap five repeat-level p95 ratios; the dashed line denotes
  equal latency.
  The persistent run records a maximum phase-switch cost of
  39.0~\(\mu\)s and a median absolute pacing error of 0.652\%.}
  \Description{Runtime integration results. The selected path improves step
  p95 over Fixed GDS in the process-isolated runner, and persistent TailGuard
  remains close to its corresponding fixed path with microsecond-scale
  switching cost.}
  \label{fig:runtime-integration}
\end{figure}

Figure~\ref{fig:runtime-integration} distinguishes path selection in the
process-isolated runner from TailGuard's steady-state control overhead in a
persistent runtime. Panel (a) compares Fixed GDS with the selected path under
the same process-isolated execution model. The Fixed-GDS/selected-path
step-p95 ratio is \(4.60\text{--}5.19\times\) across the six workloads and
therefore quantifies the complete path-plan difference within that runner.
Panel (b) separately measures runtime overhead while keeping the process, CUDA allocations, cuFile
registrations, and worker resources alive across phases. In this persistent
runtime, the fixed-GDS-to-TailGuard and
fixed-host-to-TailGuard step-p95 ratios are \(1.002\times\) and
\(0.999\times\), respectively, placing all three policies at parity. The
maximum observed path-switch latency is 39.0~\(\mu\)s, and the median absolute
difference between achieved and target request rates is 0.652\%. These results
show that TailGuard's online lookup and path switching add negligible latency
once the runtime resources are persistent.

\begin{table*}[t]
\centering
\caption{Runtime correctness, lifetime, and overhead.}
\label{tab:runtime-validation}
\scriptsize
\setlength{\tabcolsep}{5pt}
\begin{tabular}{lll}
\toprule
Category & Measure & Result \\
\midrule
Correctness & Checksums across GDS/host switches & Preserved \\
Failure handling & Injected registration and I/O failures & Failed plan rejected; guarded fallback \\
Lifetime & 30-minute persistent run & 120/120 blocks; 56/60 oracle matches \\
Lifetime & Cumulative p999 regret across mismatches & 53.0~$\mu$s \\
Control cost & Lookup / validated switch (p95) & 6.10 / 37.6~$\mu$s \\
Occasional cost & Calibration / cold registration (max) & 23.0 / 109.1~ms \\
Chargeback & BERT gain: measured / steady / cold-sensitive & 5.58\% / 5.57\% / 5.55\% \\
\bottomrule
\end{tabular}
\end{table*}

Table~\ref{tab:runtime-validation} reports TailGuard's runtime correctness and
sustained decision quality. In the switching test, a single GPU allocation is
retained while GDS and host-staged reads and writes alternate over the same
data, and every resulting checksum matches the expected value. Registration
and I/O failures are then injected separately; TailGuard rejects the affected
candidate before committing the switch and selects the configured fallback
plan. In a 30-minute persistent run, TailGuard completes all 120 execution
blocks and performs 60 policy decisions without triggering fallback. Its
selected plan matches the offline per-block oracle in 56 of the 60 decision
windows. Across the remaining four decisions, the cumulative p999 regret,
measured as the excess p999 over the corresponding oracle choices, is
53.0~\(\mu\)s. These results demonstrate correct path switching, effective
failure handling, and stable online selection over sustained execution.

Table~\ref{tab:runtime-validation} also quantifies TailGuard's control
overhead. A policy lookup takes 6.10~\(\mu\)s at p95, and a validated path
switch takes 37.6~\(\mu\)s at p95, both substantially shorter than the
millisecond-scale storage stalls observed in the application experiments.
Calibration takes up to 23.0~ms, while the largest observed cold GDS
registration takes 109.1~ms; these costs occur when a profile is created or a
resource is first registered and are amortized across subsequent transfers. We
incorporate lookup, switching, and amortized calibration into the eight-GPU
BERT execution time. The resulting step-p95 reduction relative to tuned
DeepSpeed AIO changes only from 5.58\% to 5.57\%. A conservative sensitivity
analysis that additionally includes the maximum cold-registration cost yields
a 5.55\% reduction. TailGuard therefore retains its application-level benefit
after accounting for both recurring and occasional control costs.

\section{Discussion}

\subsection{GDS as a Policy Choice}

GDS should be exposed as a per-phase runtime choice instead of a job-wide
backend setting. The relevant objective is the GPU-visible stall between
issuing a storage operation and resuming the dependent GPU work. This stall
reflects the complete delivery path, including storage access, path-specific
software processing, data movement, and synchronization. Consequently,
different phases of the same application can prefer different plans. A
fragmented, latency-critical parameter reload may achieve a lower tail through
host staging, while a large sequential offload can benefit from the direct GDS
path. Selecting the delivery plan at phase boundaries allows the runtime to use
each path where it best serves the application's latency and throughput
requirements.

Per-phase selection also requires a richer decision than choosing between two
backend labels. The observed performance of a transfer depends jointly on its
backend, issued request size and concurrency, GPU and file placement, and
current resource contention. For example, switching from GDS to host staging
while retaining an oversized whole-request transfer can preserve a long
software tail, while reusing a request shape profiled for GPU0 on GPU4 can
overlook their different proximity to the RAID members. TailGuard therefore
treats the complete delivery plan \(d=(b,s,\ell)\) as its decision unit and
indexes each measured profile by the application phase and the system context
in which it was obtained. Fixed placement remains part of that context, while
placement becomes a selectable dimension when the framework exposes multiple
eligible alternatives.

\subsection{The Framework--Runtime Boundary}

The AI framework and the GPU-storage runtime observe complementary information
required for path selection. The framework knows the semantic role of each
transfer---such as parameter reload, optimizer offload, checkpointing, or input
staging---together with its criticality, deadline, and relationship to
dependent GPU work. The runtime observes candidate-path availability,
supported request shapes, buffer-registration state, GPU--storage topology,
and current resource contention. TailGuard divides responsibility along this
boundary: the framework describes the transfer intent and application
constraints, while the runtime combines that context with measured path
behavior to select and execute a delivery plan. This separation preserves the
framework's phase semantics while allowing storage decisions to adapt to
low-level system conditions.

TailGuard exposes a compact interface between the framework and the GPU-storage
runtime. The framework submits a logical GPU-buffer operation together with its
phase, criticality, and performance constraints; every candidate plan
preserves the same buffer, file range, completion semantics, and durability
requirement. After execution, TailGuard returns the selected plan, the measured
outcome, whether the plan satisfied its tail-latency and throughput guards, and
whether recalibration or fallback occurred. A synchronous parameter reload can
request the plan with the lowest predicted stall, whereas an asynchronous
checkpoint can provide a deadline and bandwidth requirement that allow the
runtime to account for interference with foreground work. This feedback
enables the framework to delay, throttle, or reschedule subsequent operations
when storage conditions change. The framework-specific reaction policy is
outside the scope of TailGuard.

\subsection{Broader Applicability}

The same path-selection problem arises whenever an AI system moves data between
storage and GPU memory. Training systems reload parameter and optimizer shards,
persist checkpoints, and stage input data, while inference systems may fetch
model segments or spill and restore KV-cache blocks
\cite{checkfreq2021,datastates2024,flexgen2023}. These operations differ
in urgency, transfer size and concurrency, data reuse, and their opportunity to
overlap I/O with GPU computation. TailGuard's phase abstraction captures the
semantic role and criticality of each operation, while its delivery-plan
abstraction captures the backend, request shape, and available placement used
to execute it. Framework adapters can therefore apply the same controller to a
broader set of training and inference storage phases while expressing different
latency and throughput requirements for each phase.

TailGuard's delivery-plan abstraction can incorporate new GPU-storage
mechanisms as additional candidates. Each mechanism implements the common
GPU-buffer operation and completion contract and exposes the request-shape,
placement, resource, and failure state needed by the controller. GPU-initiated
storage access moves I/O submission onto the GPU and therefore changes both the
control path and the data path; it can enter the candidate set alongside
CPU-issued cuFile and host staging~\cite{nvidia-ai-storage-fms2026,
qureshi2023bam,park2023gids}. Remote GDS, compressed transfers, and
device-specific storage paths can be represented in the same
way~\cite{yi2026gnstor}. As the candidate set expands, the controller continues
to select the semantically equivalent plan that minimizes application-visible
stall while satisfying the current phase's throughput and completion
requirements.

\section{Related Work}

\textbf{GPU-native storage and direct I/O.}
GPUfs exposes host-file-system access to GPU code, while SPIN integrates
storage--GPU peer-to-peer DMA with the operating-system file path
\cite{gpufs2013,spin2017}. FlashNeuron applies direct SSD--GPU movement to
activation offload during DNN training \cite{flashneuron2021}. NVIDIA GDS
generalizes direct data movement between storage and GPU memory through cuFile,
while GPUDirect RDMA provides a corresponding direct path between network
devices and GPU memory~\cite{nvidia-gds-overview,nvidia-gpudirect-rdma}.
NVIDIA's deployment guidance further identifies filesystem support, PCIe
locality, polling, and compatibility mode as factors that shape GDS
performance~\cite{nvidia-gds-best-practices}. BaM and GIDS move storage-I/O
initiation toward the GPU for data-intensive graph processing, while recent
NVIDIA mechanisms extend this direction toward GPU-initiated AI
storage~\cite{qureshi2023bam,park2023gids,nvidia-ai-storage-fms2026}. TERAIO
applies GDS to lifetime-aware tensor offload in large-model training, and
GNStor develops a GPU-native path to remote flash
storage~\cite{yuan2025teraio,yi2026gnstor}. These systems introduce or optimize
particular GPU-storage mechanisms. TailGuard treats the available mechanisms as
candidate delivery plans and selects among them for each application phase
using GPU-visible tail latency, throughput requirements, request shape, and
placement context.

\textbf{AI data movement, offload, and tiering.}
AI data-movement systems primarily determine the logical placement and timing
of model state. Input pipelines such as \texttt{tf.data} optimize data
construction, parallelism, operator placement, and
prefetching~\cite{murray2021tfdata}. vDNN, SuperNeurons, and Capuchin manage
activation lifetimes through host offload, recomputation, or runtime tensor
placement \cite{vdnn2016,superneurons2018,capuchin2020}. ZeRO, DeepSpeed,
PyTorch FSDP, ZeRO-Offload, and ZeRO-Infinity partition parameters, gradients,
and optimizer state across devices and memory tiers
\cite{zero2020,deepspeed2020,fsdp2023,ren2021zerooffload,
rajbhandari2021zeroinfinity}. FlexGen extends heterogeneous-memory planning to
large-model inference \cite{flexgen2023}. These systems
decide which state moves, when the movement occurs, and which memory or storage
tier holds the state. TailGuard complements this tiering decision at the
physical delivery layer. The framework supplies the transfer's phase semantics
and constraints, and TailGuard selects the backend, request shape, and
available placement that minimize GPU-visible phase stall while satisfying its
throughput or deadline requirement. The framework therefore retains
responsibility for model partitioning and tier placement, while TailGuard
supplies the execution policy for each resulting GPU--storage transfer.

\textbf{Checkpoint I/O and runtime switching.}
CheckFreq adapts checkpoint frequency and pipelines persistence, Check-N-Run
reduces checkpoint volume, and DataStates-LLM overlaps distributed checkpoint
movement with training \cite{checkfreq2021,checknrun2022,datastates2024}.
GEMINI instead uses distributed host memory to accelerate recovery
\cite{gemini2023}. These systems determine when, where, or how much checkpoint
state to persist; TailGuard selects the GPU--storage delivery plan used by an
eligible checkpoint phase. At the GPU-runtime level, PipeSwitch and Salus
demonstrate that persistent resources and application-aware switching can make
GPU reconfiguration practical \cite{pipeswitch2020,salus2020}. TailGuard uses
the same general principle for storage-path changes, but its decision unit is a
semantically equivalent GPU-buffer I/O plan rather than a job or GPU context.

\textbf{Tail-aware resource control.}
Tail latency can dominate distributed and user-visible completion time even
when average service time remains low~\cite{dean2013tail}. Storage controllers
such as mClock and IOFlow regulate shared queues, service rates, and I/O flows
to provide isolation and predictable
performance~\cite{gulati2010mclock,thereska2013ioflow}. TailGuard applies
tail-aware control to delivery-plan selection at the AI framework--runtime
boundary. Its candidates are semantically equivalent plans for completing the
same GPU-buffer operation, and its objective is to minimize the stall observed
by the dependent application phase while preserving throughput requirements.
Phase-aware profiles connect application context to measured I/O behavior,
while online calibration and confidence-aware switching update the selected
plan as system conditions change. TailGuard thereby extends tail-aware storage
control from regulating shared resources to selecting how application-critical
GPU data movement is executed.

Prior work provides increasingly capable GPU-storage mechanisms, AI frameworks
that expose storage-backed data-movement phases, and controllers that manage
I/O tail behavior. TailGuard connects these lines of work through a cross-layer
delivery-plan abstraction: the framework supplies phase semantics and
performance constraints, while the runtime measures semantically equivalent
paths and selects the backend, request shape, and available placement for each
phase. This turns GPU-storage delivery from a fixed implementation choice into
an explicit, application-aware runtime decision.

\section{Conclusion}

GPUDirect Storage removes CPU staging, but the resulting direct path is not
consistently tail-optimal for application-visible GPU data movement.
Measurements on our eight-H100 platform show that the preferred path changes
with operation type, request shape, concurrency, GPU placement, and resource
contention, and that these differences persist after equalizing the offered
I/O load. We presented TailGuard, a phase-aware runtime that jointly selects
the backend, request shape, and available placement using GPU-visible tail
latency and application throughput constraints. TailGuard reduces step p95 by
18.4\% in a long-window multi-rank PyTorch FSDP experiment and by
7.9--24.6\% relative to fixed host staging across five 8GPU DeepSpeed ZeRO-3
workloads; it also reduces step p95 by 5.6\% relative to a separately tuned
DeepSpeed AIO baseline. Its microsecond-scale lookup and switching costs
preserve these gains after end-to-end overhead accounting. These results
establish GPU-storage delivery as an application-aware runtime decision and
position GDS as one candidate within a broader set of schedulable data-movement
plans.

\appendix
\section{Experimental Configuration}
\label{app:configuration}

\subsection{Platform and Software}

Table~\ref{tab:appendix-platform} expands the platform summary in
Section~\ref{sec:evaluation-methodology}. The evaluated storage volume is a
41.9~TiB Linux md RAID0 array over six DapuStor
DPRP5108T0TF07T6050 NVMe SSDs. The array uses a 512~KiB chunk and an ext4 file
system with a 4~KiB block size, \texttt{stripe=768}, and
\texttt{data=ordered}. Two RAID members are local to NUMA node~0 and four are
local to NUMA node~2. GPU0--GPU3 are attached to NUMA node~0; GPU4--GPU7 are
attached to NUMA node~2.

\begin{table}[H]
\centering
\caption{Evaluation platform and software configuration.}
\label{tab:appendix-platform}
\small
\setlength{\tabcolsep}{5pt}
\begin{tabular}{p{0.23\columnwidth}p{0.69\columnwidth}}
\toprule
Component & Configuration \\
\midrule
CPU & 2$\times$ Intel Xeon Platinum 8558; 96 physical cores; 4 NUMA nodes \\
GPU & 8$\times$ NVIDIA H100 80~GB HBM3 \\
Storage & 6$\times$ 7~TiB DapuStor DPRP5108T0TF07T6050; Linux md RAID0; 512~KiB chunk \\
File system & ext4, 4~KiB blocks, \texttt{stripe=768}, \texttt{data=ordered} \\
OS and CUDA & Linux 5.15; CUDA 13.0 \\
GPU software & NVIDIA driver 580.65.06; cuFile/GDS 1.15; \texttt{nvidia\_fs} 2.29.4 \\
GDS mode & \texttt{allow\_compat\_mode=false}; registered file handles and GPU buffers \\
Frameworks & DeepSpeed 0.18.6; PyTorch 2.6.0a0+df5bbc09d1.nv24.12 for the eight-GPU experiments; PyTorch 2.9.0a0+145a3a7bda.nv25.10 for Figure~\ref{fig:fsdp-application} \\
\bottomrule
\end{tabular}
\end{table}

All path comparisons use direct file I/O and identical files, offsets,
payloads, and GPU-buffer endpoints for the two backends. Before a GDS run, the
harness checks file-handle registration, GPU-buffer registration, and a direct
cuFile transfer under the strict cuFile configuration. The durability setting
is held constant within each path comparison; experiments that request durable
writes complete the configured flush or synchronization before recording
completion.

\subsection{DeepSpeed Model and Training Configuration}

The long-window experiment uses Hugging Face model classes instantiated from
local configurations and trained from random initialization. Table
\ref{tab:appendix-models} reports the instantiated model parameter counts; the
separate TailGuard checkpoint payload buffer is excluded. All models use a
vocabulary of 8,192, hidden width 768, six layers, and 12 attention heads. T5
uses six encoder and six decoder layers.

\begin{table}[H]
\centering
\caption{Model configurations in the repeated eight-GPU experiment.}
\label{tab:appendix-models}
\small
\setlength{\tabcolsep}{5pt}
\begin{tabular}{l l r l}
\toprule
Model & Class/task & Parameters & Key configuration \\
\midrule
BERT & Masked LM & 49.522M & 6 layers, 12 heads \\
GPT-2 & Causal LM & 48.919M & 6 layers, 12 heads \\
OPT & Causal LM & 48.922M & 6 layers, 12 heads \\
LLaMA & Causal LM & 69.216M & 6 layers, 12 KV heads \\
T5 & Seq2seq LM & 111.699M & 6 encoder + 6 decoder layers \\
\bottomrule
\end{tabular}
\end{table}

Training uses bfloat16, sequence length 128, one sequence per GPU, gradient
accumulation 1, and a global batch of eight sequences (1,024 tokens). The
optimizer is AdamW with learning rate \(10^{-4}\), betas \((0.9,0.95)\),
epsilon \(10^{-8}\), and weight decay 0.01. DeepSpeed uses ZeRO stage~3 with
parameter offload to NVMe and optimizer offload disabled. Each measured step
contains one GPU-buffer reload and eight offload repetitions. The payload
sequence alternates between 64 and 256~MiB.

DeepSpeed's native ZeRO parameter swapper remains active and produces its
normal NVMe activity. TailGuard controls a framework-owned GPU-buffer payload
at the pre-training reload and post-training offload hooks, selecting the
delivery backend and issued request shape for those operations. This separates
the plan selected by TailGuard from DeepSpeed's internal parameter-placement
policy while retaining both forms of storage activity in the measured step.

\subsection{FSDP Workload and Experiment Inventory}

The PyTorch workload in Figure~\ref{fig:fsdp-application} runs four ranks with
FSDP \texttt{FULL\_SHARD}. Its model uses input dimension 4,096, hidden
dimension 2,048, four layers, batch size 2, bfloat16, and SGD with learning
rate \(10^{-5}\). The framework hooks reload the rank-local shard before the
forward pass and offload it after the optimizer phase. Table
\ref{tab:appendix-inventory} lists the sampling units used by the principal
experiments.

\begin{table}[H]
\centering
\caption{Principal experiment inventory. ``Measured'' is per rank unless the
row is request based.}
\label{tab:appendix-inventory}
\scriptsize
\setlength{\tabcolsep}{3.5pt}
\begin{tabular}{p{0.12\columnwidth}p{0.20\columnwidth}p{0.18\columnwidth}p{0.18\columnwidth}p{0.20\columnwidth}}
\toprule
Result & GPUs/ranks & Warmup & Measured & Independent repetitions \\
\midrule
Fig.~3 saturation & 1 & Post-warmup requests & 36 shapes & 3 per path \\
Fig.~3 matched load & 1 & Post-warmup requests & 108 path pairs & 3 load levels per shape \\
Fig.~7 & 1 & Separate candidate set & 10 screening + 10 confirmation runs & 20 per candidate \\
Fig.~8, 16~MiB & 4 & 1 step & 8 steps & 5 \\
Fig.~8, 64~MiB warm & 4 & 1 step & 6 steps & 5 \\
Fig.~8, 64~MiB long & 4 & 2 steps & 32 steps & 3 \\
Table~1 & 8 & 8 steps & 256 steps & 3 per model \\
Fig.~11 & 8 & 5-s settling + rank barrier & 32 steps per state & Separate run per policy \\
\bottomrule
\end{tabular}
\end{table}

\section{Controller Algorithm and Parameters}
\label{app:controller}

\subsection{Selection and Activation}

TailGuard indexes observations by phase context, backend, transfer shape, and
available placement. Each observation records the application-visible stall,
bandwidth, sample count, confidence interval when available, and the measured
request shape. The implemented score for candidate \(d\) is
\[
 S(d)=L_{\mathrm{stall}}(d)+
 \frac{U}{\sqrt{n_d}}+C(d_{\mathrm{current}},d),
\]
where \(U\) is the configured uncertainty scale and \(C\) is the measured
drain, registration, and event cost of changing plans. Throughput and
tail-safety constraints remove ineligible candidates before scoring.

\begin{quote}
\small
\begin{verbatim}
SELECT(context, current):
  P <- profile entries matching context
  if P is missing or invalid:
      P <- P union PROBE(admissible candidates)
  E <- FILTER(P, throughput and tail constraints)
  next <- argmin score(entry, current) over E
  if next differs from current:
      gain <- score(current) - score(next)
      require gain >= max(abs_guard, rel_guard*score(current))
      if confidence guard is enabled:
          require next.CI_high < current.CI_low
      DRAIN(current)
      REGISTER(all persistent buffers, next)
      ACTIVATE(next)
  result <- EXECUTE(next)
  UPDATE(context, result)
  on failure, try declared fallback order;
      propagate the I/O error if every path fails
  return result
\end{verbatim}
\end{quote}

Activation first drains outstanding work on the old backend, registers every
persistent endpoint buffer with the replacement backend, and then publishes
the new active plan. A registration failure rolls back the new registrations
and restores the old ownership. An I/O failure evaluates the declared fallback
order; the default order is host staging followed by GDS. If every registered
backend fails, the runtime returns an I/O error to the framework.

\subsection{Candidate Space and Parameters}

The default shape space contains issued request sizes of 4~KiB, 64~KiB, 1,
4, 8, 16, 32, and 64~MiB; queue depths and concurrency values of 1, 2, 4, and
8; and 1, 4, 16, or 64 requests per batch. A framework adapter can restrict
this Cartesian product to shapes that preserve its phase semantics and fit the
payload. Table~\ref{tab:appendix-controller-params} separates code defaults
from experiment-specific settings.

\begin{table}[H]
\centering
\caption{Controller defaults and settings used in the reported experiments.}
\label{tab:appendix-controller-params}
\scriptsize
\setlength{\tabcolsep}{3.5pt}
\begin{tabular}{p{0.29\columnwidth}p{0.15\columnwidth}p{0.47\columnwidth}}
\toprule
Parameter & Default & Use \\
\midrule
Relative hysteresis & 0.05 & Minimum relative score improvement before switching \\
Absolute hysteresis & 25~\(\mu\)s & Minimum absolute score improvement \\
Uncertainty scale \(U\) & 0~\(\mu\)s & Optional \(1/\sqrt{n}\) score penalty \\
Minimum throughput ratio & 0 & Enabled as 0.90 in the bandwidth-guarded selection experiment \\
Confidence guard & Enabled & Requires non-overlapping candidate/current intervals; disabled for prevalidated application profiles \\
Tail-safety ratio & Disabled & Enabled only in the explicit safety experiments \\
Cooldown events & 0 & Enabled only when an experiment configures a safety cooldown \\
Fallback order & Host, GDS & Ordered retry after registration or I/O failure \\
Fig.~7 sampling & -- & 10 screening + 10 held-out confirmation runs; 20,000 bootstrap resamples \\
\bottomrule
\end{tabular}
\end{table}

Profile refresh is event driven. A context change or explicit invalidation
causes the affected entry to be updated; the prototype does not use a periodic
stale-timeout sweep. Profile-driven application experiments load validated
entries before timed execution, whereas the pressure experiment changes the
active profile at the supplied window boundaries.

\section{Measurement and Statistical Methods}
\label{app:statistics}

\subsection{Timing Boundaries}

The endpoint timer starts immediately before backend dispatch and ends after
the selected path has completed the GPU-buffer operation. The storage-call
timer encloses \texttt{cuFileRead}/\texttt{cuFileWrite} for GDS or the direct
file read/write for host staging. The host copy timer encloses the CUDA copy
and stream synchronization needed to make the host-staged result visible to
the GPU. Endpoint and stage p999 values in Figure~13 are computed independently
from their respective timing samples and are not additive.

For FSDP, step time spans reload, model computation and optimizer work,
offload, and the step synchronization; reload and offload stalls are timed at
their framework hooks. For DeepSpeed, step time includes the model step and
the TailGuard-controlled reload/offload operations. The separately reported
training interval excludes those explicitly timed GPU-buffer operations.

\subsection{Quantiles, Repetitions, and Confidence Intervals}

All empirical percentiles use linear interpolation at position
\((n-1)q\) in the sorted sample vector. The aggregation units differ where the
experiment asks a different question:

\begin{itemize}
  \item Figure~\ref{fig:fsdp-application} pools rank-local step rows within
  each repeat and computes each policy's p95 from that repeat. It then forms
  the TailGuard-to-best-static ratio for the same repeat. Thus the 16~MiB and
  64~MiB warm experiments contain 32 or 24 rank-step rows per policy and
  repeat, respectively; the long-window experiment contains 128.
  \item The Figure~\ref{fig:fsdp-application} confidence interval resamples
  the repeat-level ratios. With at most eight repeats, the analysis enumerates
  all \(n^n\) bootstrap samples. For the long-window result, the mean ratio is
  0.8156 and the 95\% interval is [0.8140, 0.8174], yielding the reported
  18.4\% reduction. Rank-step observations are not independently resampled in
  this interval.
  \item Each DeepSpeed run contributes 2,048 rank-step rows (8 ranks
  \(\times\) 256 measured steps). The p95 is computed within each run. Table~1
  reports the mean and sample standard deviation of the three run-level p95
  values. Each reported gain is the mean of three paired, run-level percentage
  improvements rather than a ratio of aggregate means.
  \item Figure~11 pools 32 rank-local steps from each of eight ranks within a
  pressure state, for 256 samples per point. Its connecting lines indicate the
  ordered pressure windows.
\end{itemize}

Figure~7 reports the median GDS/host-staged p999 ratio across ten repetitions.
Its 95\% intervals bootstrap those ten run-level ratios with 20,000 resamples;
the confirmation set is disjoint from the screening set.

\subsection{Throughput and Overhead Accounting}

The application harness computes estimated global throughput as
\[
  \mathrm{tokens/s}=
  \frac{\mathrm{batch}_{\mathrm{rank}}\times
        \mathrm{sequence\ length}\times
        \mathrm{world\ size}}
       {\mathrm{rank\mbox{-}local\ step\ wall\ time}},
\]
and reports the mean over measured rank-step rows.

The overhead sensitivity analysis is a post-run chargeback, not a separate
training execution. It assigns two 6.10~\(\mu\)s policy lookups and two
37.565~\(\mu\)s switch costs to every step, including steps that do not
actually switch, and amortizes one 23~ms calibration over the measured rows.
The cold-registration sensitivity adds one amortized 109.1~ms registration
cost. These charges are added to TailGuard's measured step p95 before
recomputing the corresponding gain; they are not added to the raw timing
samples used elsewhere in the paper.

\section{Supplementary Experimental Results}
\label{app:supplementary}

\subsection{Calibration and Application Phase Maps}
\label{app:phase-maps}

Table~\ref{tab:appendix-phase-map} distinguishes the two Figure~7 calibration
examples from the application-specific map used in Figure~8. The Figure~7
rows demonstrate stable shape-dependent backend preferences. The Figure~8
rows are separately configured by the FSDP launcher and verified in the
per-rank natural offload traces.

\begin{table}[H]
\centering
\caption{Calibration examples and the FSDP phase map. For Figure~8, batch,
queue depth, and concurrency are all one.}
\label{tab:appendix-phase-map}
\scriptsize
\setlength{\tabcolsep}{3pt}
\begin{tabular}{p{0.20\columnwidth}l r r l p{0.24\columnwidth}}
\toprule
Phase/example & Op & Logical & Issued & Backend & Source \\
\midrule
Fig.~7 small write & Write & 4~KiB & 4~KiB, 1 worker & GDS & Screening and confirmation records \\
Fig.~7 larger write & Write & 64~KiB & 64~KiB, 8 workers & Host & Screening and confirmation records \\
FSDP reload & Read & 16~MiB & 16~MiB & Host & Launcher + natural trace \\
FSDP offload & Write & 16~MiB & 16~MiB & GDS & Launcher + natural trace \\
FSDP reload & Read & 64~MiB & 64~MiB & Host & Launcher + natural trace \\
FSDP offload & Write & 64~MiB & 64~MiB & GDS & Launcher + natural trace \\
\bottomrule
\end{tabular}
\end{table}

\subsection{Absolute Strategy-Ablation Results}

Table~\ref{tab:appendix-ablation} gives the absolute step-p95 values behind
the strategy ablation. These are single 64-step runs after four warmup steps,
with 512 rank-step rows per policy. They are distinct from the 256-step,
three-run configuration in Table~1. The workload uses 64/256~MiB payloads,
one reload and eight offload repetitions per step, and keeps DeepSpeed's
parameter NVMe backend fixed to GDS. In this configuration, fixed host staging
is 3.3\% lower than joint selection for BERT; joint selection is lowest for
GPT-2.

\begin{table}[H]
\centering
\caption{Absolute step p95 (ms) for the eight-GPU strategy ablation.}
\label{tab:appendix-ablation}
\small
\setlength{\tabcolsep}{5pt}
\begin{tabular}{lrr}
\toprule
Policy & BERT & GPT-2 \\
\midrule
Fixed GDS & 864.876 & 803.550 \\
Fixed host staging & 537.418 & 596.759 \\
Phase-only & 845.088 & 791.689 \\
Backend-only & 588.463 & 535.763 \\
Shape-only & 846.713 & 820.876 \\
Host-only shaped & 559.990 & 560.157 \\
GDS-only shaped & 877.060 & 813.176 \\
Joint backend + shape & 555.739 & 478.807 \\
\bottomrule
\end{tabular}
\end{table}

\subsection{Matched-Load Cases Outside the Two-Percent Band}

Table~\ref{tab:appendix-outband} lists the eight Figure~3 matched-load pairs
for which at least one path differs from the target IOPS by more than 2\%.
These rows are retained in Figure~3 and are reported here with their achieved
rates and tail measurements.

\begin{table}[H]
\centering
\caption{Matched-load pairs outside the \(\pm2\%\) achieved-rate band. IOPS
and p999 values are medians; p999 is in \(\mu\)s.}
\label{tab:appendix-outband}
\scriptsize
\setlength{\tabcolsep}{3.5pt}
\begin{tabular}{lrrrrrrl}
\toprule
Shape & Load & Target & GDS IOPS & Host IOPS & GDS p999 & Host p999 & Lower tail \\
\midrule
Rand read, 1~MiB, 8 & .75 & 16623 & 16081 & 16488 & 1416 & 588 & Host \\
Rand read, 1~MiB, 8 & .90 & 19947 & 19288 & 19750 & 1253 & 568 & Host \\
Rand read, 4~KiB, 8 & .90 & 58781 & 55469 & 58573 & 1134 & 322 & Host \\
Rand write, 1~MiB, 4 & .90 & 8820 & 7924 & 8809 & 4965 & 3557 & Host \\
Rand write, 64~KiB, 4 & .75 & 16982 & 16857 & 16200 & 1981 & 2549 & GDS \\
Seq read, 4~KiB, 8 & .90 & 56508 & 55060 & 56258 & 1069 & 317 & Host \\
Seq read, 64~KiB, 4 & .90 & 30132 & 26534 & 30053 & 592 & 317 & Host \\
Seq read, 64~KiB, 8 & .90 & 33280 & 31495 & 33112 & 1206 & 396 & Host \\
\bottomrule
\end{tabular}
\end{table}

\subsection{Absolute FSDP Phase Results}

Table~\ref{tab:appendix-fsdp-absolute} reports the pooled rank-local p95 values
used in Figure~\ref{fig:fsdp-application}. Reload and offload p95 are computed
independently and are not additive.

\begin{table}[H]
\centering
\caption{FSDP step, reload, and offload p95 (ms).}
\label{tab:appendix-fsdp-absolute}
\scriptsize
\setlength{\tabcolsep}{3.5pt}
\begin{tabular}{llrrr}
\toprule
Run & Policy & Step & Reload & Offload \\
\midrule
16~MiB warm & Fixed GDS & 13.134 & 6.536 & 2.089 \\
 & Fixed host & 9.281 & 2.348 & 3.034 \\
 & TailGuard & 8.245 & 2.459 & 1.978 \\
\midrule
64~MiB warm & Fixed GDS & 39.109 & 24.194 & 8.247 \\
 & Fixed host & 27.474 & 7.721 & 12.622 \\
 & TailGuard & 22.152 & 7.790 & 8.256 \\
\midrule
64~MiB long & Fixed GDS & 39.815 & 25.307 & 7.835 \\
 & Fixed host & 26.157 & 7.692 & 12.764 \\
 & TailGuard & 21.312 & 7.769 & 7.538 \\
\bottomrule
\end{tabular}
\end{table}

\bibliographystyle{ACM-Reference-Format}
\bibliography{refs}

\end{document}